\documentclass[pdflatex,sn-mathphys-num]{sn-jnl-bettermargins}

\usepackage{graphicx}%
\usepackage{multirow}%
\usepackage{amsmath,amssymb,amsfonts}%
\usepackage{amsthm}%
\usepackage{mathrsfs}%
\usepackage[title]{appendix}%
\usepackage{xcolor}%
\usepackage{textcomp}%
\usepackage{manyfoot}%
\usepackage{booktabs}%
\usepackage{algorithm}%
\usepackage{algorithmicx}%
\usepackage{algpseudocode}%
\usepackage{listings}%
\usepackage{enumitem}
\usepackage{tabularx}
\newcolumntype{L}{>{$}l<{$}} 
\newcolumntype{R}{>{$}r<{$}} 
\newcolumntype{K}{>{$}c<{$}} 
\newcolumntype{C}[1]{>{\centering\arraybackslash}p{#1}} 
\newcommand{\Rzero}{\ensuremath{\mathcal{R}_0}}
\newcommand{\vecalpha}{\ensuremath{\vec{\alpha}}}
\newcommand{\boldA}{\ensuremath{\boldsymbol{A}}}

\theoremstyle{thmstyleone}%
\newtheorem{theorem}{Theorem}
\newtheorem{proposition}{Proposition}%
\newtheorem{corollary}{Corollary}%

\theoremstyle{thmstyletwo}%

\theoremstyle{thmstylethree}%

\begin{document}
\title[Once bitten, twice shy: A modeling framework for incorporating heterogeneous mosquito biting into transmission models]{Once bitten, twice shy: A modeling framework for incorporating heterogeneous mosquito biting into transmission models}


\author*[1,2,3]{\fnm{Kyle J.-M.} \sur{Dahlin}}\email{kyledahlin@vt.edu}

\author[1,2,3]{\fnm{Michael A.} \sur{Robert}}

\author[1,2,3]{\fnm{Lauren M.} \sur{Childs}}

\affil[1]{\orgdiv{Department of Mathematics}, \orgname{Virginia Tech}, \orgaddress{\street{225 Stanger Street}, \city{Blacksburg}, \postcode{24061}, \state{Virginia}, \country{USA}}}

\affil[2]{\orgdiv{Center for the Mathematics of Biosystems}, \orgname{Virginia Tech}, \orgaddress{\street{225 Stanger Street}, \city{Blacksburg}, \postcode{24061}, \state{Virginia}, \country{USA}}}

\affil[3]{\orgdiv{Center for Emerging, Zoonotic, and Arthropod-borne Pathogens}, \orgname{Virginia Tech}, \orgaddress{\street{1015 Life Science Circle}, \city{Blacksburg}, \postcode{24061}, \state{Virginia}, \country{USA}}}


\abstract{The risk and intensity of mosquito-borne disease outbreaks are tightly linked to the frequency at which mosquitoes feed on blood, also known as the biting rate. 
However, standard models of mosquito-borne disease transmission inherently assume that mosquitoes bite only once per reproductive cycle---an assumption commonly violated in nature. 
Drivers of multiple biting, such as host defensive behaviors or climate factors, also affect the mosquito gonotrophic cycle duration (GCD), a quantity customarily used to estimate the biting rate. 
Here, we present a novel framework for incorporating more complex mosquito biting behaviors into transmission models. 
This framework can account for heterogeneity in and linkages between mosquito biting rates and multiple biting. 
We provide general formulas for the basic offspring number, $\mathcal{N}_0$, and basic reproduction number, \Rzero{}, threshold measures for mosquito population and pathogen transmission persistence, respectively. 
To exhibit its flexibility, we expand on specific models derived from the framework that arise from empirical, phenomenological, or mechanistic modeling perspectives. 
Using the gonotrophic cycle duration as a standard quantity to make comparisons among the models, we show that assumptions about the biting process strongly affect the relationship between GCD and \Rzero{}. 
While under the standard assumption of one bite per reproductive cycle, \Rzero{} is an increasing linear function of the inverse of the GCD, alternative models of the biting process can exhibit saturating or concave relationships. 
Critically, from a mechanistic perspective, decreases in the GCD can lead to substantial decreases in \Rzero{}. 
Through sensitivity analysis of the mechanistic model, we determine that parameters related to probing and ingesting success are the most important targets for disease control. 
This work highlights the importance of incorporating the behavioral dynamics of mosquitoes into transmission models and provides a method for evaluating how individual-level interventions against mosquito biting scale up to determine population-level mosquito-borne disease risk.}

\keywords{mosquito-borne disease, biting dynamics, disease control, phase-type distributions, generalized linear chain trick}



\maketitle
\section{Introduction}\label{sec1}

The burden and geographic spread of mosquito-borne diseases, such as dengue, have risen substantially over the past decade \citep{Malavige2023-sn}, and outbreaks of newly-emerged zoonotic viruses like Zika virus are occurring at increasing rates \citep{Evans2018-ip}. 
Often, the actual wildlife reservoirs of such zoonoses remain unidentified, posing significant challenges to the prevention or suppression of spillover \citep{Haydon2002-cc}. 
Concurrently, malaria continues to claim hundreds of thousands of lives every year, primarily children under five \citep{World-Health-Organization2022-jh}. 
While efforts to suppress malaria led to consistent success in the early part of the 21st century, progress has stalled, particularly since the start of the global COVID-19 pandemic \citep{World-Health-Organization2022-jh}. 
For mosquito-borne diseases, the difficulty in reducing cases of endemic disease, suppressing epidemics, or preventing the emergence of zoonoses is tied to the mosquito's incredible effectiveness at locating and feeding on hosts. 

Unlike direct or environmental transmission, vector-borne disease transmission is not incidental. 
Pathogens exploit the fluid exchange intrinsic to mosquito blood-feeding to enable transmission to vertebrate hosts or mosquitoes. 
However, mosquito contact alone does not determine that transmission will occur. 
For example, a mosquito may land on an infectious host but not become exposed to a pathogen due to disruption prior to the start of blood ingestion or while attempting to find a blood vessel on which to feed. 
On the other hand, disruption can also enhance transmission if it occurs at later stages, after transmission has occurred but before the mosquito has reached a state of repletion, having attained a full blood meal. 
There is, therefore, a key distinction between blood-feeding---the successful ingestion of a bloodmeal from a host---and biting---an attempt or series of attempts to blood-feed, each of which may or may not be successful. 
We use the term \textit{biting process} to refer to the entire process of mosquito host-seeking, biting, and eventual blood-feeding, whether the mosquito is successful at any of these stages or not. 
Understanding how the biting process interacts with transmission is critical to the development of effective interventions to suppress endemic mosquito-borne diseases and predict the risk of the emergence or spread of novel diseases.

While the biting process is biologically and ecologically complex, it is typically represented simply in models. 
The biting process is difficult to observe in the field, and thus, data to fit relevant model parameters are lacking \citep{Silver2008-cy}. 
In the face of this challenge, modelers commonly make the assumption that the rate at which mosquitoes bite hosts (the biting rate) is equal to the inverse of the gonotrophic cycle duration (GCD), the average length of time it takes a mosquito to complete its reproductive cycle once \citep{Smith2012-fi,Zahid2023-ne}. 
However, this formulation, which we will call the \textit{standard biting rate}, can limit the ability of modelers to examine aspects of the mosquito biting process critical to mosquito population dynamics and the transmission dynamics of mosquito-borne disease. 

We suggest that a vicious cycle may drive the mismatch between the complexity of the biting process and the simplicity of its representation in models. 
Because the biting process is represented simply in most models and its effect on transmission is straightforward, empiricists may see no need to investigate and measure characteristics of the biting process more deeply. 
Consequently, modelers could view the lack of data on the biting process as evidence of the simplicity of the process and, thus, see no need to produce more sophisticated models. 
Despite this, several modelers have developed models incorporating more realistic biting dynamics \citep{Anderson1999-mt,Teboh-Ewungkem2021-wo,Ghakanyuy2022-hn}. 
We aim to break this cycle by developing a generalized framework of the mosquito biting process, which will enable the creation of models designed to achieve different goals or utilize diverse sources and types of data. 

Essential to the formulation of the standard biting rate is the additional assumption that a mosquito takes a single blood meal in each gonotrophic cycle, an assumption commonly violated in nature \citep{Scott2012-ac}. 
Modifying the standard biting rate assumption to account for multiple biting is ordinarily done straightforwardly: the biting rate is multiplied by the average number of bites per gonotrophic cycle or the \textit{multiple biting number}. 
Thus, if a mosquito completes its gonotrophic cycle in $T$ days (i.e., its GCD is $T$) and bites $M$ times during that cycle, the modified biting rate is now $M/T$ instead of $1/T$. 
However, this modification includes a crucial implicit assumption, namely, that the biting rate and the multiple biting number are independent---a strong assumption that may not be realistic. 
Consider, for example, the effect of disruption due to host defensiveness on the biting process.
Disruption increases the GCD (thereby decreasing $1/T$), but it also increases the number of bites the mosquito takes in a single cycle (increasing $M$). 
The ultimate effect of disruption on biting rate depends on the relative changes in the GCD and multiple biting number. 
Biotic factors, such as larval conditions, body size, or infection status, and abiotic factors, such as temperature or rainfall, are critical determinants of mosquito biting frequency that may also create a link between the GCD and multiple biting number \citep{Thavara2001-uz,Delatte2009-gq,Wei-Xiang2022-kh,Zahid2023-ne}. 

The standard assumptions used to formulate biting rates induce a direct relationship between multiple biting and transmission: transmission potential (measured through the basic reproduction number, $\mathcal{R}_0$) is directly correlated with the multiple biting number \citep{Tedrow2019-gy}. 
However, correlations between the GCD and the multiple biting number could substantially alter this effect. 
As a simple example, if some factor increases the multiple biting number from $M$ to $aM$ and the GCD from $T$ to $aT$, the modified biting rate, $aM/aT = M/T$, is unchanged.
Contrary to the usual expectation, in this case, increasing the multiple biting number will not lead to an increase in transmission potential. 
In light of renewed calls to study the effects of heterogeneity in mosquito biting on transmission, there is a need for flexible models that can represent the interrelationship between multiple biting and biting rates \citep{Blanken2024-yn}. 

In this study, we develop a generalized modeling framework for the mosquito biting process and incorporate it into a transmission model. 
Despite the additional complexity introduced in this framework, we show how quantities commonly used to determine population persistence (the basic offspring number, $\mathcal{N}_0$) and transmission potential (the basic reproduction number, $\mathcal{R}_0$) can still be calculated through straightforward formulas. 
We then exhibit how this general framework can be specified to represent the mosquito biting process empirically (using measurements of mosquito gonotrophic cycle durations), phenomenologically (by reproducing observed patterns), or mechanistically (by specifying relevant transition rates and probabilities directly). 
In standard models \Rzero{} is directly correlated to the standard biting rate (i.e., the inverse of the GCD). 
By calculating \Rzero{} as a function of this standard biting rate, we show how these different framework specifications lead to distinct predictions about the relationship between GCD and transmission potential. 
Finally, focusing on the mechanistic model, we determine which rates or probabilities associated with biting most strongly impact the basic offspring number and \Rzero{}.

\subsection{The mosquito biting process}\label{sec:biting_process}
The mosquito biting process broadly comprises four steps: host-seeking, landing, probing, and ingesting \citep{Edman1988-am}. 
If a mosquito succeeds at each of these steps, it will obtain the blood necessary for egg development and oviposition. 
But there are many ways in which a mosquito's attempt to complete these steps may be thwarted, leading to a need to re-land on a host or quest for a new one. 
Each feeding disruption event might lead to more opportunities for transmitting pathogens between vertebrate hosts and mosquitoes, thereby increasing the overall potential for outbreaks in the system \citep{Edman1971-ot}. 
Below, we briefly describe the steps of the biting process and discuss their possible effects on transmission. 

Most mosquito species that are capable of transmitting disease require a blood meal from a host to develop eggs \citep{Edman1988-am}. 
Newly emerged adult female mosquitoes and those that have recently laid eggs (oviposited) must first successfully locate the host upon which it will feed, a process called host-seeking. 
Mosquitoes follow active or passive approaches to host-seeking \citep{Edman1988-am}. 
Those mosquitoes employing an active approach go on flights to areas where hosts are more likely to be encountered. 
Whereas, passive mosquitoes lie in wait until a potential blood meal source is detected. 
Both methods require mosquitoes to synthesize multiple cues to locate a host \citep{Carde2015-rf,Wynne2020-cl}. 
The probability of locating a host is not only a function of the homing ability of the mosquito but also the availability of hosts in the area, which is itself determined by the population density of hosts, the preference of the mosquito for that host (or similarly, the host's attractiveness), and the tolerance of the host to mosquito biting \citep{Agusto2013-rw, Blanken2024-yn}. 

Once a suitable host is located, the mosquito must successfully land on the host. 
The mosquito, if detected, may have its attempt thwarted depending on the ability and propensity of the host to engage in defensive behavior. 
If the attempt is thwarted, the mosquito may attempt to land again on the same host or return to the searching phase to seek out a new one. 
After successfully landing, the mosquito often waits to ensure it has not been detected, then begins foraging for a site on the host's body to probe for blood \cite{Edman2020-fh}. 

The mosquito then probes for a blood vessel from which it can ingest blood. 
At this stage, mosquito saliva is distributed in the skin to promote aggregation of blood in the area \citep{Edman1988-am}, and the transmission of infectious agents from the mosquito to the vertebrate host is most likely to occur \cite{Thongsripong2021-hv}. 
If the mosquito is detected, defensive behaviors by the host might dislodge the mosquito (necessitating a new attempt at landing or a departure to locate a new host) or otherwise increase the probing duration or the number of probing attempts. 
Infected mosquitoes tend to probe longer than uninfected ones \citep{Edman1988-am}. 

After successfully finding a vein, the mosquito begins ingesting blood. 
To initiate egg development, the mosquito requires a critical amount of blood determined by its body size, current nutritional reserves, and the type of host \citep{Kelly2001-ac,Farjana2013-it}. 
If a mosquito is dislodged before ingesting this critical amount of blood, it may attempt to land on another part of the same host or depart to locate a new host. 
Pathogen transmission from infectious vertebrate hosts to mosquitoes can occur during ingestion, even if mosquitoes do not acquire a complete blood meal \citep{Thongsripong2021-hv}. 
However, transmission from mosquitoes to hosts at this stage is unlikely, due to the high speed of blood flow towards the mosquito's mouthparts \cite{Edman2020-fh}. 
If the mosquito has recently taken a partial blood meal, the ingestion process might occur more quickly as it requires a smaller volume of blood. 
Additionally, this successive feeding increases the likelihood that the mosquito acquires a infection by producing avenues for pathogen infection through micro-perforations in the mosquito gut \citep{Armstrong2020-st,Brackney2021-sv}. 

Once the mosquito attains a supply of blood sufficient for egg development, it enters the oviposition state. 
In this state, the engorged mosquito is at an exceptionally high risk of attack by predators or defense by the host. 
It thus immediately leaves the hosts to seek refuge while digesting the blood meal \citep{Roitberg2003-rk}. 
The blood meal is digested over two to four days \cite{Edman2020-fh}. 
Egg deposition generally occurs one to two days after digestion is completed. 
The rate of this process is highly dependent on temperature and varies across species \citep{Oda1980-ha,Alto2001-nq,Yang2009-fi,Moser2023-ih}. 
After egg deposition, the mosquito rests for some time and usually feeds on plant sugar sources before returning to biting \cite{Edman2020-fh}. 

Altogether, the process of blood-feeding, oviposition, resting, and back to biting is called the gonotrophic cycle. 
While biting forms the critical link to transmission, through the actions of probing and ingesting, other parts of the cycle also contribute to the overall transmission cycle because they strongly determine the time between successive bites and mosquito mortality. 

\subsection{Modeling processes with phase-type distributed waiting times}\label{sec:phase-type}
An important aspect of our model framework is the representation of the waiting time distribution (also known as the dwell-time distribution) between when a mosquito first enters and then exits the biting state. 
We provide a summary of some important properties of phase-type distributions here, but we direct readers to \cite{Hurtado2019-st, Hurtado2021-dz} for concrete descriptions of how they can be used in ecological and epidemiological modeling and to \cite{Bladt2017-gp} for a comprehensive reference on matrix-exponential distributions. 
Phase-type distributions, written $\operatorname{PH}(\vec{\alpha}, \boldsymbol{A})$, are a subclass of matrix exponential distributions parameterized by an initial probability vector $\vec{\alpha}$ (note that we will follow the convention that all vectors are column vectors) and a transient rate matrix $\boldsymbol{A}$. 
For any discrete or continuous-time Markov chain with a set of transient states $\vec{X} = \left[X_1,\ldots, X_n\right]^T$, the dwell-time distribution, the distribution describing the time between entering and exiting the transient states, is represented by a phase-type distribution (though this representation is not unique). 
Here, the initial probability vector $\vec{\alpha}=\left[\alpha_1,\ldots,\alpha_n\right]^T$ defines how the inflow of new entrants into the system is distributed among the transient states, namely, the probability that a new individual enters the transient states at state $X_i$ is given by $\alpha_i$. 
The transient rate matrix $\boldsymbol{A}$ includes all the information on the transition rates among the transient states as well as the overall loss rates from each of the transient states to the absorbing state(s), given by $-\boldsymbol{A}\vec{1}$. 
A non-diagonal entry of \boldA{}, $a_{ij}>0$, $i\neq j$, gives the transition rate from transient state $X_i$ to transient state $X_j$. 
After specifying the net rates out of each transient state to the absorbing states (i.e., the entries of the vector $-\boldsymbol{A}\vec{1}$), the diagonal entries of \boldA{}, $a_{ii}<0$, are determined. 
If $v_i$ is the net rate out of transient state $X_i$ into the absorbing state(s), then $a_{ii} = -v_i - \sum_{j=1}^{n} a_{ij}$ . 
Through the Generalized Linear Chain Trick (Corollary 2, \cite{Hurtado2019-st}), the mean-field dynamics of a Markov process with this associated phase-type representation are described by a system of ordinary differential equations.
\begin{equation}\label{eq:phase_example}
    \frac{d}{dt}\vec{X} = \vec{\alpha}\mathcal{I}(t) + \boldsymbol{A}^T\vec{X},
\end{equation}
where $\mathcal{I}(t)$ is the total inflow rate into the transient states $\vec{X}$.

Similar to exponentially-distributed dwell-times, phase-type distributed dwell-times exhibit relationships between their mean and rate parameters that aid their interpretation. 
The matrix $U = \left(-\boldsymbol{A}\right)^{-1}$, also called the \textit{Green matrix} of $\operatorname{PH}(\vec{\alpha}, \boldsymbol{A})$, has the property that its entries $u_{ij}$ give the expected amount of time the process spends in transient state $X_j$ before absorption, given that it initially entered through state $X_i$ (Theorem 3.1.14, \cite{Bladt2017-gp}). 
Note that the non-singularity of $\boldsymbol{A}$ is guaranteed as long as the $X_i$ are all transient states (Theorem 3.1.11, \cite{Bladt2017-gp}). 
In the case of phase-type distributions, the mean dwell-time duration is $\vec{\alpha}^T\left(-\boldsymbol{A}\right)^{-1}\vec{1}$. 
The eigenvalues of $\boldsymbol{A}$ have strictly negative real parts (Theorem 3.1.15, \cite{Bladt2017-gp}), a property that will be helpful in later equilibrium analyses. 

Finding a phase-type distribution approximating the dwell-time distribution of any given process is possible because phase-type distributions are dense in the space of positive distributions (Theorem 3.2.9, \cite{Bladt2017-gp}). 
In practice, one can estimate $\vec{\alpha}$ and $\boldsymbol{A}$ from empirical data through likelihood expectation-maximization algorithms, many of which have been implemented in MATLAB, R, and Python \citep{Horvath2012-ht, Asmussen1996-vp}. 
Alternatively, one can select any positive distribution for their dwell-time (possibly also fit to data) and find a corresponding phase-type distribution through moment matching or Monte-Carlo methods \citep{Horvath2012-ht, Bladt2017-gp}. 
Finally, if the underlying Markov chain is specified through some mechanistic understanding of the process being modeled (i.e. the transient state transition diagram is known), then $\vec{\alpha}$ and $\boldsymbol{A}$ can be determined directly from the entrance and exit rates from each of the transient states and the probabilities of transition between them \citep{Hurtado2021-dz}. 

\section{Model formulation and analysis}
We begin by developing a generalized framework for integrating the mosquito biting process into a model for mosquito population dynamics, then discuss the existence and stability of its equilibria in terms of the basic offspring number, $\mathcal{N}_0$. 
This framework is then assimilated into a model for disease transmission between mosquitoes and vertebrate hosts. 
We calculate the basic reproduction number, \Rzero{}, and show that it is a threshold quantity for disease outbreaks and persistence.

\subsection{Demographic model}\label{sec:model-form}

The mosquito population is divided into four subpopulations representing its life history and reproductive states: the aquatic state compartment containing all types of juvenile mosquitoes (eggs, larvae, and pupae), $J$; the biting state compartment(s), $B$; the oviposition state compartment, $V$; and the post-oviposition resting state compartment, $R$.
$B$ includes all mosquitoes requiring a blood meal to oviposit, either newly-emerged mosquitoes or those mosquitoes returning from oviposition after resting. 
$V$ includes all mosquitoes that are actively ovipositing. 
$R$ includes all mosquitoes resting after ovipositing. 
$J$, $B$, $V$, and $R$ are referred to as states, while any of their sub-compartments are called stages. 

Focusing on the biting population, $B$, we assume that the distribution for the time spent in the biting state is a phase-type distribution, $\operatorname{PH}(\vecalpha{}, \boldA{})$. 
Suppose that there are $n$ transient stages for the biting state so that $B$ can be decomposed as a vector $\vec{B}=\left[B_1,B_2,\cdots,B_{n-1},B_n\right]^T$ of biting stages $B_i$.
Then $\vec{\alpha}$ is an $n\times1$ vector whose entries, $\alpha_i$, represent the proportion of mosquitoes newly entering biting stage $i$, or equivalently, the probability that an individual mosquito enters the biting state at stage $i$. 
The transient rate matrix, $\boldsymbol{A}$, is an $n\times n$ matrix whose off-diagonal entries, $a_{ij}$ $(i\neq j)$, represent the instantaneous rates of transition to biting stage $i$ from biting stage $j$. 
Let $t_j$ denote the rate that a mosquito in biting stage $j$ exits the biting state. 
Then the diagonal entries of $\boldsymbol{A}$ are given by $a_{jj} = t_j - \sum_{i\neq j} a_{ij}$ where $a_{jj} < 0$, to ensure that state $j$ is in fact a transient state and not an absorbing state. 

\begin{table}[t!]
    \centering
    \caption{State variables for system \eqref{eq:orig_model}.}
    \label{tab:orig_state_variables}
    \begin{tabular}{Kl}
        \hline
        \textrm{Symbol} & Description \\
        \hline
         J & Juvenile mosquitoes (eggs, larvae, and pupae) \\
         \vec{B} & Biting-state mosquitoes  \\
         V & Ovipositing mosquitoes  \\
         R & Resting mosquitoes  \\\hline
    \end{tabular}
\end{table}

\begin{figure}[t!]
        \centering
        \includegraphics[width=0.75\textwidth,trim={0 2.1in 3in 0},clip]{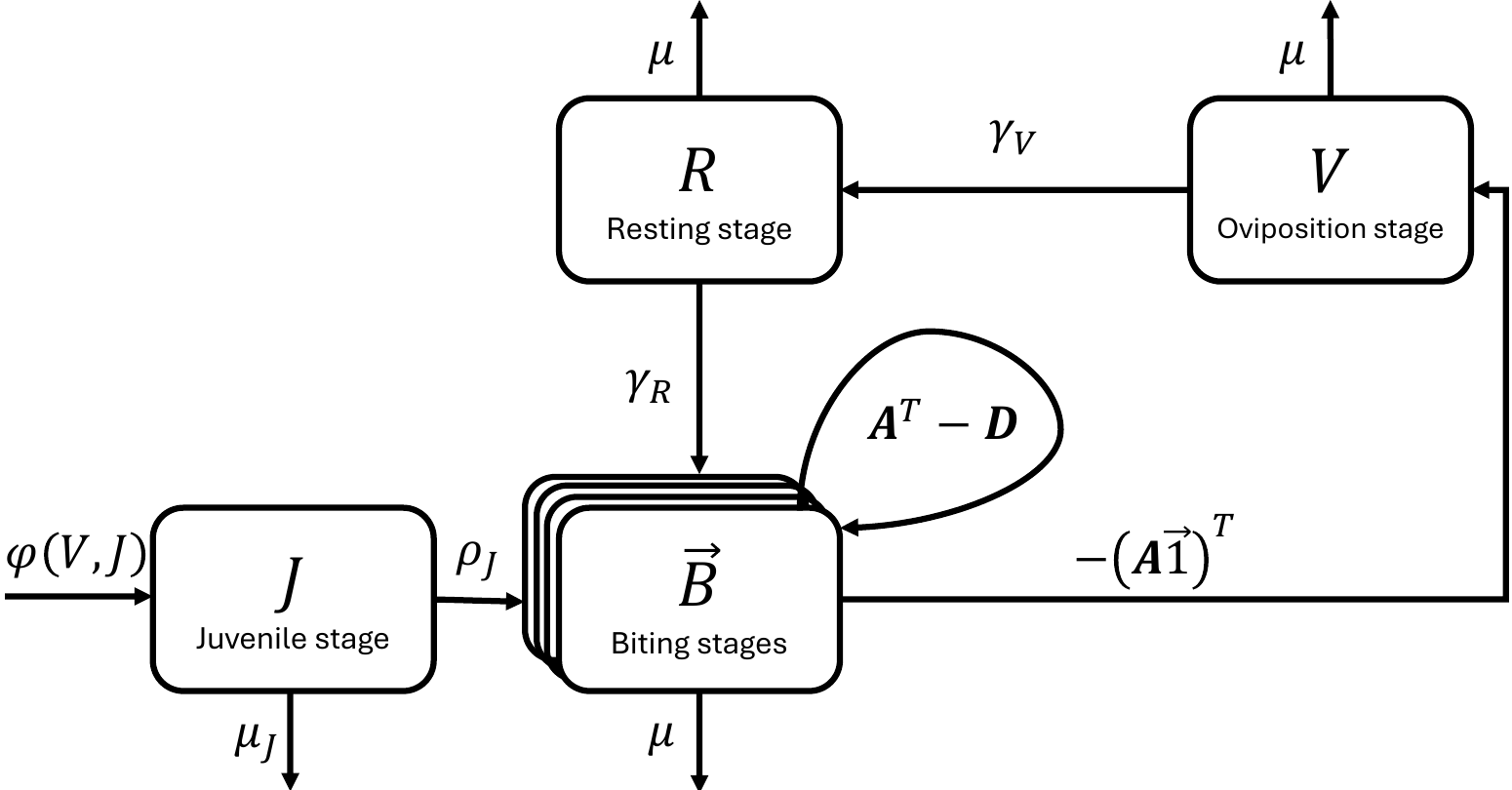}
        \caption{Compartmental diagram for the mosquito demographic model, system \eqref{eq:orig_model}. The matrix $\boldsymbol{A}^T -\boldsymbol{D}$ indicates the transition rates within and among the biting stages. $\boldsymbol{D}=\boldsymbol{D}({\operatorname{diag}\boldsymbol{A}-(\boldsymbol{A\vec{1})^T}})$ is the diagonal matrix whose entries are given by the vector $\operatorname{diag}\boldsymbol{A}-(\boldsymbol{A\vec{1})^T}$, where $\operatorname{diag}\boldsymbol{A}$ is itself the column vector whose entries are the diagonal entries of $\boldsymbol{A}$.}\label{fig:general_pop_diagram}
\end{figure}

We assume simple forms for the recruitment and mortality rates for each stage, as our focus is on biting dynamics. 
Aquatic-stage mosquitoes are recruited at the recruitment rate $\varphi(V, J)$, develop into adults at the rate $\rho_J$, and endure density-independent mortality at the per-capita rate $\mu_J$. 
Biting, ovipositing, and resting state mosquitoes experience natural mortality at the per-capita rate $\mu$. 
Finally, the transition rates from oviposition to resting and from resting to biting are $\gamma_V$ and $\gamma_R$, respectively. 
By the Generalized Linear Chain Trick (Corollary 2, \cite{Hurtado2019-st}), the mean field model of this process is:
\begin{equation}\label{eq:orig_model}
\begin{aligned}
\frac{d}{dt}J&=\varphi(V,J)- \rho_{J}J -\mu_{J}J,\\
\frac{d}{dt}\vec{B}&=\left(\rho_{J}J+\gamma_R R\right)\vec{\alpha}+\boldsymbol{A}^{T}\vec{B}-\mu\vec{B},\\
\frac{d}{dt}V&=\left(-\boldsymbol{A}\vec{1}\right)^{T}\vec{B}-\gamma_V V-\mu V,\\
\frac{d}{dt}R&=\gamma_V V - \gamma_R R -\mu R.
\end{aligned}
\end{equation}
Table \ref{tab:orig_state_variables} defines the state variables, and Table \ref{tab:parameters} provides the complete list of parameters and their descriptions. 
Figure \ref{fig:general_pop_diagram} shows the compartmental diagram associated with the system.

\begin{table}[t!]
    \caption{Descriptions of the parameters of systems
    \eqref{eq:orig_model}, \eqref{eq:epi_model}, and baseline values used for producing simulated data.}
    \label{tab:parameters}
\centering
\def\arraystretch{1.1}
\begin{tabular}{cp{0.5\linewidth}p{0.175\linewidth}}
\hline
Parameter & Description & Value \tabularnewline
\hline
\multicolumn{3}{c}{Mosquito life history parameters}\tabularnewline
\hline
$\mu$ & Adult mosquito mortality rate & $1/\left(20 \text{ days}\right)$ \tabularnewline
$\gamma_V$ & Transition rate from ovipositing to resting & $1/\left(5 \text{ days}\right)$ \tabularnewline
$\gamma_R$ & Transition rate from resting to host-seeking & $1/\left(2 \text{ days}\right)$ \tabularnewline
$K_J$ & Juvenile mosquito carrying capacity & $2250$ individuals \tabularnewline
$\rho_J$ & Juvenile mosquito development rate & $1/\left(12 \text{ days}\right)$ \tabularnewline
$\mu_J$ & Juvenile mosquito mortality rate & $1/\left(20 \text{ days}\right)$ \tabularnewline
$\varphi$ & Eggs per female per day & $3$ \tabularnewline
\hline
\multicolumn{3}{c}{Epidemiological parameters}\tabularnewline
\hline
$\beta_H$ & Probability of transmission from vector to host & 50\% \tabularnewline
$\beta_B$ & Probability of transmission from host to vector & 50\% \tabularnewline
$\eta$ & Parasite development rate in the mosquito & $1/\left(7 \text{ days}\right)$ \tabularnewline
$\gamma_H$ & Host recovery rate & $1/\left(5 \text{ days}\right)$ \tabularnewline
\hline
\multicolumn{3}{c}{Host life history parameters}\tabularnewline
\hline
$\mu_H$ & Host mortality rate & $1/\left(65 \text{ years}\right)$ \tabularnewline
$K_H$ & Host carrying capacity & $3000$ individuals\tabularnewline
\end{tabular}
\end{table}

With this formulation, the average GCD, the time it takes a mosquito to complete a single reproductive cycle, is given by 
\begin{align}
    \textrm{GCD} &= \vec{\alpha}^{T}\left( -\boldsymbol{A}\right)^{-1}\vec{1} + \frac{1}{\gamma_V} + \frac{1}{\gamma_R}\label{eq:GCD}.
\end{align}
In this formula for the GCD, we have neglected mortality, $\mu$, to focus solely on the dynamics of the gonotrophic cycle. 
Because the average durations of the oviposition stage ($1/\gamma_V$) and resting stage ($1/\gamma_R$) are independent of the parameterization of the biting process, we focus on the first term of \eqref{eq:GCD}, the average time spent in the biting stages,
\begin{align}\label{eq:theta_dfn}
    \theta = \vec{\alpha}^{T}\left( -\boldsymbol{A}\right)^{-1}\vec{1}.
\end{align}

For system \eqref{eq:orig_model} to have at least one stable positive equilibrium, the recruitment rate function for juvenile mosquitoes, $\varphi(V,J)$, must be a smooth function of the ovipositing and aquatic-stage mosquito population sizes, $V$ and $J$. 
Recruitment should not occur if there are no ovipositing mosquitoes, but should be positive even when $J$ is small. 
Finally, there must be some value of $J$ so that the recruitment rate is proportional, with a given constant, to the ovipositing mosquito population. 
In summary, the function $\varphi(V,J)$ must meet the following assumptions:
\begin{enumerate}[leftmargin = 0.5in, label={$J \arabic*.$}, ref=\arabic*,noitemsep,nolistsep]
    \item $\varphi:\mathbb{R}_{\ge0}^{2}\to\mathbb{R}$ and $\varphi$ is differentiable in $V$ and $J$;
    \item $\varphi$ is proportional to $V$ with $\varphi\left(0,J\right)=0$, and $\lim_{J\to0^{+}}\frac{\partial}{\partial V}\varphi\left(J,V\right)>0$;
    \item For a given $c\in\mathbb{R}_{>0}$, there exists at least one $J^*\in\mathbb{R}_{>0}$ such that \mbox{$\varphi\left(V, J^*\right) = c V$}.
\end{enumerate}

In later analysis, we use the function \mbox{$\varphi(V,J) = \phi V \left(1-\frac{J}{K_J}\right)$}, which represents a density-dependent recruitment rate with $K_J$ corresponding to the maximum population of juveniles where the recruitment rate is positive, akin to a carrying capacity. 
Another commonly-used function that satisfies these properties is $\varphi\left(V,J\right)=\phi V$, where $\phi$ is a constant per capita recruitment rate. 
However, note that assumption $J2$ precludes the use of $\varphi\left(V,J\right)=c$, where here $c$ is a constant recruitment rate. 

\subsubsection{Existence and stability of equilibria}
System \eqref{eq:orig_model} always admits a trivial equilibrium $\left(J,\vec{B},V, R\right) = \left(0,\vec{0},0,0\right)$ which we term the extinction equilibrium. 
The existence of a positive equilibrium for system \eqref{eq:orig_model} now relies on the form of the recruitment rate equation $\varphi(V,J)$.

Before describing the conditions for the existence and stability of a positive equilibrium, we first define several quantities with clear biological interpretations. 
\mbox{$\tau=\vec{\alpha}^T\left(\mu\boldsymbol{I}-\boldsymbol{A}\right)^{-1}\left(-\boldsymbol{A}\vec{1}\right)$} is the total probability of surviving and exiting the biting stages. 
$\tau$ can be viewed as the ratio of two sets of rates: the exit rates from the biting states to the oviposition state, $-\boldsymbol{A}\vec{1}$, and all of the transition rates within and out of the biting states including mortality, $\mu \boldsymbol{I}-\boldsymbol{A}$. 
In analogy with exponential distributions, $\tau$ represents the probability that a mosquito exits the biting state before succumbing to mortality. 
The term $\frac{\gamma_{R}}{\mu+\gamma_{R}}\frac{\gamma_{V}}{\mu+\gamma_{V}}$ gives the probability that an adult mosquito survives both the oviposition and resting states and, thus, $\tau\frac{\gamma_{R}}{\mu+\gamma_{R}}\frac{\gamma_{V}}{\mu+\gamma_{V}}$ is the total probability that an adult mosquito completes one entire gonotrophic cycle. 
Hence, we define \mbox{$n_G = \left(1-\tau\frac{\gamma_{R}}{\mu+\gamma_{R}}\frac{\gamma_{V}}{\mu+\gamma_{V}}\right)^{-1} = \sum_{i=1}^{\infty} \left(\tau\frac{\gamma_{R}}{\mu+\gamma_{R}}\frac{\gamma_{V}}{\mu+\gamma_{V}}\right)^{i}$}, which is the average number of gonotrophic cycles survived by an adult mosquito. 

\begin{proposition}\label{thm:base_eq_exist}
Suppose that $\varphi(V,J)$ satisfies assumptions $J1$-$J3$. Let $\bar\varphi(J) = \varphi(V,J)/V$ and define \mbox{$\tau=\vec{\alpha}^T\left(\mu\boldsymbol{I}-\boldsymbol{A}\right)^{-1}\left(-\boldsymbol{A}\vec{1}\right)$}, \mbox{$n_G = \left(1-\tau\frac{\gamma_{R}}{\mu+\gamma_{R}}\frac{\gamma_{V}}{\mu+\gamma_{V}}\right)^{-1}$}, and \mbox{$\varrho = \left[\tau\frac{1}{\mu+\gamma_{V}}\left(\frac{\rho_{L}}{\rho_{L}+\mu_{L}}\right)n_G\right]^{-1}$}. 
For each $x\in \bar\varphi^{-1}(\varrho)$, there exists a positive equilibrium for system \eqref{eq:orig_model}.
\end{proposition}
\begin{proof}
    See Appendix \ref{proof:positive_equilibria}. The set $\bar\varphi^{-1}(\varrho)$ is non-empty because of assumption $J3$ above.     
    $-\left(\mu \boldsymbol{I}-\boldsymbol{A}\right)$ is invertible because it is the transient rate matrix for the distribution given by the minimum of an exponential distribution with parameter $-\mu$ and the phase-type distribution with parameters $\vec\alpha$ and $\boldsymbol{A}$, which is itself a phase-type distribution (Corollary 3.1.32 in \cite{Bladt2017-gp}). 
\end{proof}

\begin{theorem}
    Define $\tau=\vec{\alpha}^T\left(\mu\boldsymbol{I}-\boldsymbol{A}\right)^{-1}\left(-\boldsymbol{A}\vec{1}\right)$ as in Proposition \ref{thm:base_eq_exist} and let $\mathcal{N}_0$, the basic offspring number, be defined as $$\mathcal{N}_0=\phi_{V}\tau\frac{1}{\mu+\gamma_V}\left(\frac{\rho_{J}}{\left(\rho_{J}+\mu_{J}\right)-\phi_{J}}\right)n_G$$ where $\phi_{J}=\frac{\partial\varphi}{\partial J}\left(0,0\right)$ and $\phi_{V}=\frac{\partial\varphi}{\partial V}\left(0,0\right)$.  The extinction equilibrium is locally asymptotically stable if and only if $\mathcal{N}_0<1$. 
\end{theorem}
\begin{proof}
    See Appendix \ref{proof:stability_extinction}.
\end{proof}

We term $\mathcal{N}_0$ the basic offspring number of the mosquito population because the mosquito population persists if and only if $\mathcal{N}_0 > 1$. 
The basic offspring number can be decomposed into a product of biologically meaningful quantities as
\begin{align*}
\mathcal{N}_0 =&\hphantom{a}\tau\times \frac{\phi_{V}}{\mu+\gamma_V}\times \left(\frac{\rho_{J}}{\rho_{J}+\mu_{J}-\phi_{J}}\right)\times n_G,
\end{align*}
the product of the probability of completing blood-feeding to reach oviposition ($\tau$), the average number of eggs laid during oviposition, the probability that an immature mosquito survives to adulthood, and the average number of gonotrophic cycles. 

\begin{corollary}
    If $\varphi(V,J) = \phi \left(1-\frac{J}{K_J}\right)V$, then there is a unique positive equilibrium, and the extinction equilibrium is unstable if and only if $\mathcal{N}_0 > 1$.
\end{corollary}
\begin{proof}
    In this case, $\mathcal{N}_0 = \phi/\varrho$ and the equation $\phi \left(1-\frac{J}{K_J}\right) = \phi/\mathcal{N}_0$ is solved for $J>0$ to determine the existence of a positive equilibrium.
\end{proof}

\begin{corollary}
    If $\varphi(V,J) = \phi V$, then there is a unique positive equilibrium if and only if $\mathcal{N}_0 = 1$, in which case the extinction equilibrium is unstable.
\end{corollary}

\subsection{Demographic model specification}
In system \eqref{eq:orig_model}, the biting process is totally characterized by the vector \vecalpha{} and the transient rate matrix \boldA{}. 
We now show how \vecalpha{} and \boldA{} can be determined through the lens of three different modeling philosophies---empirical, phenomenological, or mechanistic---reflecting the various goals a modeler might have for simulating a system.

\paragraph{Empirical framework}
$\vec{\alpha}$ and $\boldsymbol{A}$ can be estimated directly from lab or field measurements of the dwell-time of mosquitoes in the biting state, that is, the duration of time between a mosquito ovipositing and returning to the oviposition stage, using likelihood expectation-maximization algorithms \citep{Horvath2012-ht, Asmussen1996-vp}. 
However, because phase-type distributions are necessarily over-parameterized, there will be infinitely many pairs of $\vec{\alpha}$ and $\boldsymbol{A}$ that can accurately represent the same observed distribution \citep{Bladt2017-gp}. 
Therefore, it is necessary to specify a dimension or number of transient states before attempting to fit to data. 
In practice, one can perform the fitting process for models of increasing dimensions, $n$, then use a model selection criterion like Akaike's Information Criterion, noting that the number of parameters increases as $n^2 + n$ \citep{Akaike1998-aa}.
Alternative algorithms exist that fit phase-type distributions using the first $2n-1$ sample moments from dwell-time data to approximate $\boldsymbol{A}$ and $\vec{\alpha}$ up to dimension $n$ \citep{Horvath2017-ro}. 

\paragraph{Phenomenological framework}
If sufficient data are not available to reliably approximate $\vec{\alpha}$ and $\boldsymbol{A}$, one can structure them to replicate observed phenomena of mosquito biting and blood-feeding. 
To provide concrete examples, one might wish to replicate the distribution of the number of distinct blood meals detected in a sample of caught mosquitoes as in \cite{Scott2000-wg} or the individual heterogeneity in time between bites as in \cite{Christofferson2022-hi}. 
To exhibit the flexibility of the framework, we provide two examples here. 

First, suppose that the number of blood meals a mosquito needs to attain repletion is determined by the conditions they have endured in its juvenile stage. 
Evidence suggests mosquitoes reared in lower resource environments are more likely to require multiple blood meals to attain repletion \citep{Ramasamy2000-zr}. 
For simplicity, we assume in this example that the maximum number of blood meals needed is three, but this can be straightforwardly expanded to accommodate any number. 
Now we divide the biting stage into states, $B_i$, that indicate the number of blood meals needed by mosquitoes in that state. 
Then, these states are further divided into sub-states, $B_{i,j}$, where $j$ is the number of blood meals a mosquito has already taken, $0\le j\le i-1$. 
For instance, the state $B_{3,1}$ includes all mosquitoes that require three blood meals to attain repletion, having already received one blood meal. 
Assuming further that the rate at which a mosquito completes the biting stage, $b$, is independent of the number of blood meals it requires, we arrive at the phase-type distribution 
\begin{equation}\label{eq:fate_PH}
P_F = \operatorname{PH}\left(\vecalpha{}_F, \boldA{}_F\right)\textrm{ with }\vec{\alpha}_F = \begin{bmatrix}\rho_1\\
\rho_2\\
0\\
\rho_3\\
0\\
0
\end{bmatrix}, \boldsymbol{A_F} = \begin{bmatrix}-b & 0 & 0 & 0 & 0 & 0\\
0 & -2b & 2b & 0 & 0 & 0\\
0 & 0 & -2b & 0 & 0 & 0\\
0 & 0 & 0 & -3b & 3b & 0\\
0 & 0 & 0 & 0 & -3b & 3b\\
0 & 0 & 0 & 0 & 0 & -3b
\end{bmatrix}.
\end{equation}
The quantities $\rho_1$, $\rho_2$, and $\rho_3$ are the proportions of mosquitoes requiring one, two, or three blood meals to reach repletion, respectively, where $\rho_1+\rho_2+\rho_3 = 1$. 
One can verify that the average duration spent in the biting state across all mosquitoes, $1/b$, is equal to the average duration spent in the biting state by mosquitoes originating in $B_{1,0}$, $B_{2,0}$, or $B_{3,0}$. 

In the second example, we assume that heterogeneity in the number of blood meals taken by mosquitoes is driven by disruption due to host behavioral defenses. 
Again, for simplicity, we assume that the maximum number of blood meals that can be taken is three (i.e., the likelihood of surviving more than three attempts is vanishingly small). 
Let $\sigma_i$ denote the probability that a mosquito successfully blood feeds to repletion on its $i^{\textrm{th}}$ bite. 
Thus, we arrive at the phase-type distribution 
\begin{equation}\label{eq:disrupt_PH}
 P_D = \operatorname{PH}\left(\vecalpha{}_D, \boldA{}_D\right) \textrm{ with }\vec{\alpha}_D = \begin{bmatrix}1\\
0\\
0
\end{bmatrix},\,\boldsymbol{A}_{D}=\begin{bmatrix}-\lambda & \left(1-\sigma_{1}\right)\lambda & 0\\
0 & -\lambda & \left(1-\sigma_{2}\right)\lambda\\
0 & 0 & -\lambda
\end{bmatrix}.
\end{equation}
Assuming that the numbers of mosquitoes having taken one, two, or three bites corresponds to the equilibrium numbers of mosquitoes in $B_1^*$, $B_2^*$, $B_3^*$, respectively, one can calculate $\sigma_1$ and $\sigma_2$ from
\begin{align*}
\sigma_{1}&=\left(\frac{\lambda}{\mu+\lambda}\right)^{-1}\left(\frac{\lambda}{\mu+\lambda}-\frac{B_{2}^*}{B_{1}^*}\right)\approx 1-\frac{B_{2}^*}{B_{1}^*},\\
\sigma_{2}&=\left(\frac{\lambda}{\mu+\lambda}\right)^{-1}\left(\frac{\lambda}{\mu+\lambda}-\frac{B_{3}^*}{B_{2}^*}\right)\approx 1-\frac{B_{3}^*}{B_{2}^*}.
\end{align*}
Note that this is only valid if $B_3^* \le \left(\frac{\lambda}{\mu+\lambda}\right)B_2^*\le \left(\frac{\lambda}{\mu+\lambda}\right)^2 B_1^*$. 
In this case, the average duration spent in the biting state is $\theta = 1/\lambda+\left(1-\sigma_{1}\right)/\lambda+\left(1-\sigma_{1}\right)\left(1-\sigma_{2}\right)/\lambda$ and therefore, given $\sigma_1$ and $\sigma_2$, one can set $\lambda$ to obtain any chosen value of $\theta$.

Both $P_F$ (eq. \ref{eq:fate_PH}) and $P_D$ (eq. \ref{eq:disrupt_PH}) have the same number of parameters, can be made to have the same mean, and can be fit with the same type of data---counts of the numbers of mosquitoes biting once, twice, or three times in a single gonotrophic cycle. 
However, a modeler's perspective or research goals should determine the model used. 
As we will see in Section \ref{sec:case_study}, although $P_F$ and $P_D$ model the same phenomenon, they may produce divergent predictions regarding transmission potential.

\paragraph{Mechanistic framework}
In contrast to the phenomenological framework, knowledge of the specific mechanisms that determine the biting process, rather than just observing phenomena, can be used to formulate the system directly. 
In this case, one must explicitly define the transient states to correspond to discernible system states. 
As with models derived from exponential and gamma-distributed dwell-times, one can use the total exit rates out of each state and the transition probabilities among the states to specify \boldA{}. 
Namely, the total rate out of state $i$ is $-a_{ii}$, and the transition rate from state $i$ to state $j$ is $a_{ij}$. 
The probability of transition from state $i$ to state $j$ is then $p_{ij} = a_{ij} / a_{ii}$. 
A more detailed accounting of developing a transient rate matrix from mechanistic knowledge of a system is provided in \cite{Hurtado2019-st}. 
We provide a concrete example of a model developed from this framework in Section \ref{sec:case_study}.

\subsection{Transmission model}\label{sec:transmission}
Determining how to represent transmission requires making several critical decisions. 
To allow for flexibility in this representation, we extend system \eqref{eq:orig_model} to include a generalized form of pathogen transmission between populations of mosquitoes and vertebrate hosts. 
We make several simplifying assumptions, $T1$ - $T5$, that can be relaxed if desired, but which are fairly common across the epidemiological modeling literature. 
\begin{enumerate}[leftmargin = 0.5in, label={$T \arabic*.$}, ref=\arabic*,noitemsep,nolistsep]
    \item there is no vertical transmission of pathogen within the mosquito population;
    \item the intrinsic incubation period of pathogen development within the vertebrate host can be neglected because it is very short relative to the transmission dynamics;
    \item once infected, mosquitoes remain infected for the remainder of their lifespan;
    \item there is no infection-induced mortality in either the vertebrate host or mosquito;
    \item mosquito reproduction is not directly dependent on its contact rates with the focal host because they are generalists in their blood-feeding and seek alternative hosts if the abundance of the focal host is low.
\end{enumerate}

The population dynamics of the total host population, $H$, are governed by a simplified version of the logistic equation, \mbox{$\frac{d}{dt}H=\mu_{H}K_{H} - \mu_{H}H$}, where $K_H$ is the carrying capacity.
We divide the host population into subpopulations of susceptible ($S_{H}$), infectious ($I_{H}$), and recovered ($R_{H}$) individuals. 
We similarly divide the adult mosquito biting ($\vec{B}$), ovipositing ($V$), and resting ($R$) subpopulations into susceptible \mbox{($\vec{S}_B$, $S_V$, $S_R$)}, exposed ($\vec{E}_B$, $E_{V}$, $E_R$), and infectious ($\vec{I}_B$, $I_{V}$, $I_R$) compartments, respectively. 
Table \ref{tab:state_variables} gives the complete list of state variables. 
Throughout this section, the total number of mosquitoes in the biting state is denoted $B = \vec{1}^T\vec{B}$. 

\begin{table}[t!]
    \centering
    \caption{State variables for system \eqref{eq:epi_model}}.
    \label{tab:state_variables}
    \begin{tabular}{Kl}
        \hline
        \textrm{Symbol} & Description \\
        \hline
         \multicolumn{2}{c}{Mosquito state variables} \\\hline
         J & Juvenile mosquitoes (eggs, larvae, and pupae) \\
         \vec{S}_B & Susceptible, biting-state mosquitoes  \\
         \vec{E}_B & Exposed (infected), biting-state mosquitoes  \\
         \vec{I}_B & Infectious, biting-state mosquitoes  \\
         S_V & Susceptible, ovipositing mosquitoes  \\
         E_V & Exposed (infected), ovipositing mosquitoes  \\
         I_V & Infectious, ovipositing mosquitoes  \\
         S_R & Susceptible, resting mosquitoes  \\
         E_R & Exposed (infected), resting mosquitoes \\
         I_R & Infectious, resting mosquitoes  \\\hline
         \multicolumn{2}{c}{Host state variables} \\\hline
         S_H & Susceptible hosts  \\
         I_H & Infectious hosts  \\
         R_H & Recovered (and immune) hosts
         \\\hline
    \end{tabular}
\end{table}

The force of infection is the per-capita rate at which susceptible individuals become infected. 
Let $\varPhi_{BH}$ represent the force of infection from mosquitoes to vertebrate hosts and $\boldsymbol{\varPhi_{HB}}$ the force of infection from vertebrate hosts to mosquitoes. 
Each force of infection is the product of three independent probabilities, as shown in:
\begin{equation}
    \begin{aligned}
    \begin{tabular}{L}
        \textrm{Force of}\\
        \textrm{infection}
    \end{tabular}\hspace{-2mm} &= 
    \hspace{-2mm}  
    \begin{tabular}{K}
    \underline{\textbf{Transmission}}
    \cr\Pr\left(\begin{tabular}{L} 
    \textrm{contact leads to} \\
    \textrm{becoming infected}
    \end{tabular}
    \right) \\
    \textit{Appendix \ref{sec:transmission_probability}}
    \end{tabular}
   \hspace{-2mm}\times\hspace{-2mm}
    \begin{tabular}{K} 
    \underline{\textbf{Contact}}\\
     \Pr\left(\begin{tabular}{L}
     \textrm{contact is made} \\
     \textrm{in }\left(t,t+\delta t\right)
     \end{tabular}
     \right) \\
    \textit{Appendix \ref{sec:contact_rates}} 
    \end{tabular}
    \hspace{-2mm}\times\hspace{-2mm} 
    \begin{tabular}{K}
    \underline{\textbf{Prevalence}}
    \cr\Pr\left(\begin{tabular}{L} \textrm{contact was with}  \\
    \textrm{an infectious individual}
    \end{tabular}\right) \\
    \textit{Appendix \ref{sec:prob_contact_w_infective}}
    \end{tabular}\nonumber
    \end{aligned}
\end{equation}

Equation \eqref{eq:gen_forces_of_infection} defines both forces of infection: the force of infection from all types of mosquitoes to hosts, $\Phi_{BH}$, and the forces of infection from hosts to each type of mosquito, $\boldsymbol{\Phi_{HB}}$. 
$\boldsymbol{\beta_H}$ is a diagonal matrix whose entries are the transmission success probabilities for contact between a host and an infectious mosquito in biting stage $i$. 
Similarly, $\boldsymbol{\beta_B}$ is a diagonal matrix whose entries are the transmission success probabilities for contact between a mosquito in biting stage $i$ and an infectious host.
The contact rate matrices $\boldsymbol{\Lambda_{BH}}$ and $\boldsymbol{\Lambda_{HB}}$ encode the rates at which the different biting stage mosquitoes make contact with hosts. The exact definition of these matrices depends on how one chooses to define `contact' (discussed in detail in Appendix \ref{sec:contact_rates}). 
The vector $\vec{I}_B/B$ and scalar $I_H/H$ give the prevalence of infection in biting state mosquitoes and hosts, respectively. 
Complete details on how each part of the force of infection equations can be formulated are found in Appendix \ref{sec:specific_transmission_model}.

\begin{equation}\label{eq:gen_forces_of_infection}
\begin{aligned}
\varPhi_{BH}	&=\vec{1}^{T}\boldsymbol{\beta_{H}}\boldsymbol{\Lambda_{BH}}\frac{\vec{I}_B}{B},\\
\boldsymbol{\varPhi_{HB}}	&=\boldsymbol{\beta_{B}}\boldsymbol{\Lambda_{HB}}\frac{I_{H}}{H}.
\end{aligned}
\end{equation}
In models with a single biting stage, $B$, a commonly-used formulation of the forces of infection is obtained by setting $\boldsymbol{\beta}_H = \beta_H$, $\boldsymbol{\beta_B} = \beta_B$, $\boldsymbol{\Lambda_{HB}} = \lambda_B$, and $\boldsymbol{\Lambda_{BH}} = \frac{B}{H}\lambda_B$. 
In this case, $\Phi_{BH} = \beta_H \left(\frac{B}{H}\lambda_B \right)\frac{I_B}{B}$ and $\Phi_{HB} = \beta_B \lambda_B \frac{I_H}{H}$.

\begin{figure}[t!]
    \centering
    \includegraphics[width=\textwidth, trim={0 2.5in 0.7in 0},clip]{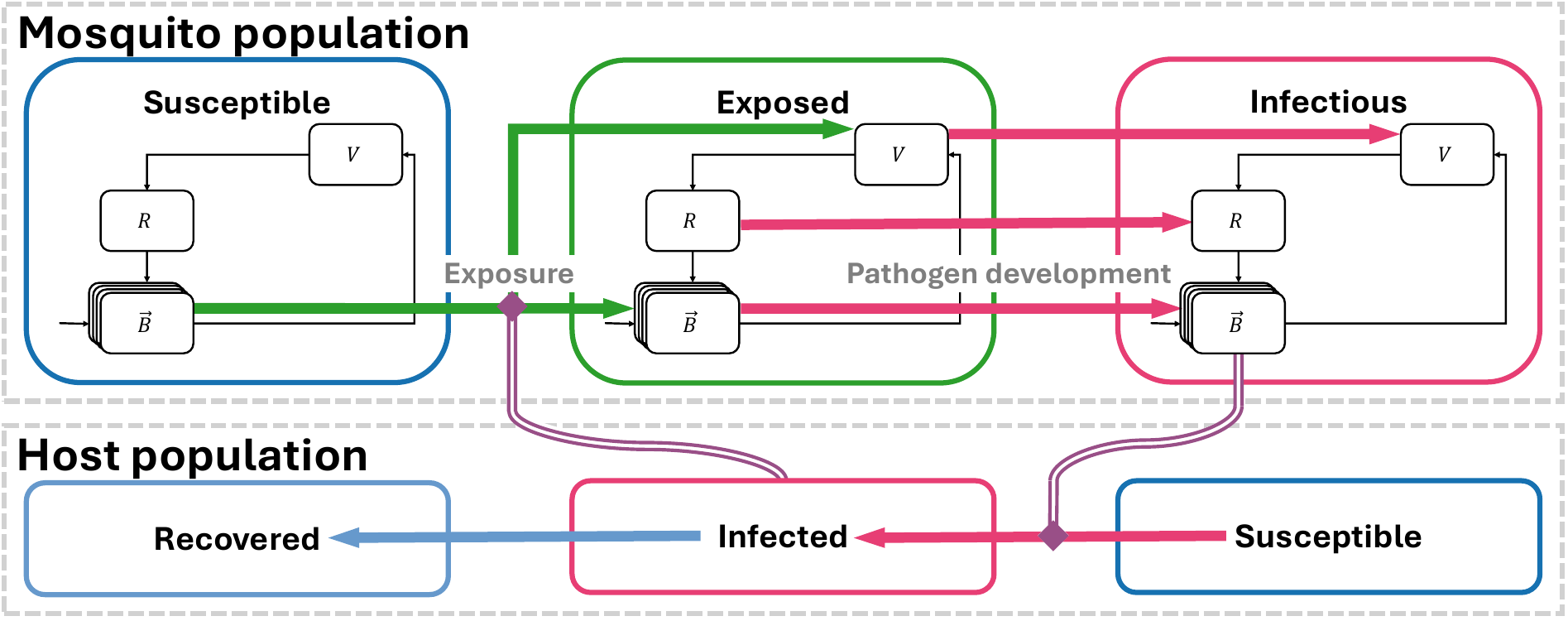}
    \caption{Compartmental diagram of the transmission model represented by system \eqref{eq:epi_model}. The oviposition and resting stages of the mosquito population, which do not contribute directly to transmission, are omitted for space considerations. Arrows indicate direct transitions between compartments. The purple double lines indicate the influence of compartments on rates of transition to infected stages, i.e., the forces of infection.}
    \label{fig:epi_diagram}
\end{figure}

With the forces of infection specified, the entire epidemiological model is expressed by:
\begin{equation}\label{eq:epi_model}
\begin{alignedat}{2}
&\frac{d}{dt}S_{H}&&=\mu_{H}K_{H}-\varPhi_{BH}S_{H}-\mu_{H}S_{H} ,\\
&\frac{d}{dt}I_{H}&&=\varPhi_{BH}S_{H}-\gamma_{H}I_{H}-\mu_{H}I_{H},\\
&\frac{d}{dt}R_{H}&&=\gamma_{H}I_{H}-\mu_{H}R_{H},\\
&\frac{d}{dt}J&&=\varphi\left(V,J\right)-\left(\rho_{J}+\mu_{J}\right)J,\\
&\frac{d}{dt}\vec{S}_B&&=\left(\rho_{J}J+\gamma_R S_R\right)\vec{\alpha}-\boldsymbol{\varPhi_{HB}}\vec{S}_B+\boldsymbol{A}^{T}\vec{S}_B-\mu\vec{S}_B,\\
&\frac{d}{dt}\vec{E}_B&&=\left(\gamma_R E_{R}\right)\vec{\alpha}+\boldsymbol{\varPhi_{HB}}\vec{S}_B-\eta\vec{E}_B+\boldsymbol{A}^{T}\vec{E}_B-\mu\vec{E}_B,\\
&\frac{d}{dt}\vec{I}_B&&=\left(\gamma_R I_{R}\right)\vec{\alpha}+\eta\vec{E}_B+\boldsymbol{A}^{T}\vec{I}_B-\mu\vec{I}_B,\\
&\frac{d}{dt}S_{V}&&=\left(-\boldsymbol{A}\vec{1}\right)^T\vec{S}_B-\gamma_V S_{V}-\mu S_{V},\\
&\frac{d}{dt}E_{V}&&=\left(-\boldsymbol{A}\vec{1}\right)^T\vec{E}_B-\eta E_{V}-\gamma_V E_{V}-\mu E_{V},\\
&\frac{d}{dt}I_{V}&&=\left(-\boldsymbol{A}\vec{1}\right)^T\vec{I}_B+\eta E_{V}-\gamma_V I_{V}-\mu I_{V},\\
&\frac{d}{dt}S_{R}&&= \gamma_V S_{V} - \gamma_R S_{R} -\mu S_{R},\\
&\frac{d}{dt}E_{R}&&= \gamma_V E_{V}-\eta E_{R} - \gamma_R E_{R}-\mu E_{R},\\
&\frac{d}{dt}I_{R}&&= \eta E_{R}+\gamma_V I_{V}-\gamma_R I_{R}-\mu I_{R}.
\end{alignedat}
\end{equation}
Table \ref{tab:parameters}, introduced previously, provides the complete list of parameters and their descriptions. 
Figure \ref{fig:epi_diagram} shows the compartmental diagram associated with the system.

\subsubsection{Existence and stability of disease-free equilibria}
Let $\left(J^{\diamond},\vec{B}^{\diamond},V^{\diamond},R^{\diamond}\right)$ be any equilibrium of system \eqref{eq:orig_model}. 
Then \mbox{$(S_{H},I_{H},R_{H},J,\vec{S}_B,\vec{E}_B,\vec{I}_B,S_{V},E_{V},I_{V},S_{R},E_{R},I_{R})=(K_{H},0,0,J^{\diamond},\vec{B}^{\diamond},\vec{0},\vec{0},V^{\diamond},0,0, R^{\diamond},0,0)$} is an equilibrium of the full system \eqref{eq:epi_model}. 
We assume that there is at least one non-extinction equilibrium of system \eqref{eq:orig_model}, which we label $(J^{*},\vec{B}^{*},V^{*},R^{*})$. 
Any equilibrium of the form $(K_{H},0,0,J^{*},\vec{B}^{*},\vec{0},\vec{0},V^{*},0,0,R^{*},0,0)$ is called a disease-free equilibrium (DFE) of system \eqref{eq:epi_model}.

Because there is no infection-induced mortality in either population, system \eqref{eq:epi_model} can be reduced by assuming that the total vector and host populations are at their disease-free equilibrium values. 
That is, \mbox{$K_{H}=S_{H}+I_{H}+R_{H}$}, \mbox{$J=J^{*}$}, \mbox{$\vec{B}^{*}=\vec{S}_B+\vec{E}_B+\vec{I}_B$}, \mbox{$V^{*}=S_V+E_{V}+I_{V}$, and $R^{*}=S_R+E_{R}+I_{R}$}. Let $B^* = \vec{1}^T\vec{B}^*$ be the total biting mosquito population size at equilibrium.


Following the next-generation matrix approach, we determine the stability of a $\textrm{DFE}$ \citep{van-den-Driessche2002-ck, Diekmann1990-yo}. To do so, we calculate the spectral radius of the next-generation matrix $\boldsymbol{K}$ for system \eqref{eq:epi_model} at the DFE and show that the DFE is stable when the spectral radius is less than one. This quantity is equivalent to the basic reproduction number, $\mathcal{R}_0$ \citep{van-den-Driessche2002-ck}. 
To derive $\boldsymbol{K}$, however, we must first specify what a new infection means in both the vertebrate host and mosquito populations. 
For our purposes, new infections in hosts are counted when individuals first enter the $I_H$ stage from $S_H$. 
In mosquitoes, new infections occur in those individuals entering any of the $\vec{E}_B$ stages directly from $\vec{S}_B$ (i.e., the rates of entry are in the vector $\boldsymbol{\varPhi_{HB}} \vec{S}_B$).

We first define a few intermediate quantities before expressing the full basic reproduction number. 
The $i,j$ entry of the non-negative matrix $\left(\frac{\gamma_{V}}{\mu+\gamma_{V}}\right)\left(\frac{\gamma_{R}}{\mu+\gamma_{R}}\right)\left(-\boldsymbol{A}\vec{1}\right)\vec{\alpha}^{T}$ gives the average rate of re-entry into biting stage $j$ after exiting $i$ (accounting for the fact that mortality may have occurred in the ovipositing or resting stages). 
Similarly, the non-negative matrix $\left(\frac{\gamma_{V}}{\mu+\gamma_{V}+\eta}\right)\left(\frac{\gamma_{R}}{\mu+\gamma_{R}+\eta}\right)\left(-\boldsymbol{A}\vec{1}\right)\vec{\alpha}^{T}$ gives re-entry rates for exposed mosquitoes, accounting for the probability that mosquitoes may have moved from exposed to infected while in the ovipositing or resting stages. 
The $i$,$j$ entry of the matrix \mbox{$\boldsymbol{\Gamma_{I}}=\left[\left(\mu\boldsymbol{I}-\boldsymbol{A}\right)-\left(\frac{\gamma_{V}}{\mu+\gamma_{V}}\right)\left(\frac{\gamma_{R}}{\mu+\gamma_{R}}\right)\left(-\boldsymbol{A}\vec{1}\right)\vec{\alpha}^{T}\right]^{-1}$} gives the average amount of time a mosquito spends in biting stage $j$ while being infectious, given that it initially became infectious in biting stage $i$. 
Similarly, the $i,j$ entry of the matrix \mbox{$\boldsymbol{\Gamma_{E}}=\left[\left(\left(\mu+\eta\right)\boldsymbol{I}-\boldsymbol{A}\right)-\left(\frac{\gamma_{V}}{\mu+\gamma_{V}+\eta}\right)\left(\frac{\gamma_{R}}{\mu+\gamma_{R}+\eta}\right)\left(-\boldsymbol{A}\vec{1}\right)\vec{\alpha}^{T}\right]^{-1}$} gives the average time a mosquito in biting stage $j$ spends being in the exposed state, given that it was initially exposed in biting stage $i$. 
The $i,j$ entry of the matrix \mbox{$\boldsymbol{\tau_{E}}=\boldsymbol{\Gamma_{E}}\left(\eta\boldsymbol{I}+\left[\left(1-\frac{\eta}{\mu+\gamma_{R}+\eta}\right)\left(\frac{\eta}{\mu+\gamma_{V}+\eta}\right)+\left(\frac{\eta}{\mu+\gamma_{R}+\eta}\right)\right]\left(\frac{\gamma_{R}}{\mu+\gamma_{R}}\right)\left(-\boldsymbol{A}\vec{1}\right)\vec{\alpha}^{T}\right)$} is therefore the probability that a mosquito that became exposed in biting stage $i$ survives to become infectious in biting stage $j$. 
Intuitively, one can think of the matrix $\boldsymbol{\tau_{E}}$ as representing the product of a matrix of durations, $\boldsymbol{\Gamma_E}$, and a matrix of rates, resulting in a matrix of probabilities.
From these matrices, we can also determine the average infectious period, $\Gamma_I = \vec{\alpha}^T \boldsymbol{\Gamma_I} \vec{1}$, the extrinsic incubation period, $\Gamma_E = \vec{\alpha}^T \boldsymbol{\Gamma}_E \vec{1}$, and the total probability of an exposed mosquito becoming infectious, $\tau_E = \vec{\alpha}^T \boldsymbol{\tau_E} \vec{1}$.

The basic reproduction number is
\begin{equation}\label{eq:R0}
\mathcal{R}_0 = \sqrt{\left(\frac{1}{\mu_{H}+\gamma_{H}}\right)\vec{1}^{T}\boldsymbol{\beta_{H}}\boldsymbol{\Lambda_{BH}}\boldsymbol{\Gamma_{I}}^T\boldsymbol{\tau_{E}}^T\boldsymbol{\beta_{B}}\boldsymbol{\Lambda_{HB}}\frac{\vec{B}^{*}}{B^{*}}}.
\end{equation}

\begin{theorem}\label{thm:DFE_stability}
    Given $\mathcal{R}_0$ in equation \eqref{eq:R0}, a \textrm{DFE} for system \eqref{eq:epi_model} is unstable if $\mathcal{R}_0>1$ and stable if \Rzero$<1$.
\end{theorem}
\begin{proof}
The proof is a straightforward application of Theorem 2 of \citet{van-den-Driessche2002-ck}. The computation of $\mathcal{R}_0$ is provided in Appendix \ref{proof:DFE_stability}.
\end{proof}

This equation can be further decomposed as vectors of mosquito-to-host and host-to-mosquito type reproduction numbers. Let $\mathcal{R}_{B_iH}$ denote the average number of new infections in hosts induced by a single infectious mosquito in biting stage $i$ in a completely susceptible population. Similarly, $\mathcal{R}_{HB_i}$ is the average number of new infections in mosquitoes in biting stage $i$ induced by a single infectious host in a completely susceptible population. 
\begin{align} \label{eq:type_R0s}
    \vec{\mathcal{R}}_{BH} = \begin{bmatrix}
        \mathcal{R}_{B_1 H} \\
        \mathcal{R}_{B_2 H} \\
        \vdots \\
        \mathcal{R}_{B_n H}
    \end{bmatrix} &= \left(\vec{1}^{T}\boldsymbol{\beta_{H}}\boldsymbol{\Lambda_{BH}}\boldsymbol{\Gamma_{I}}^T\boldsymbol{\tau_{E}}^T\right)^T, \\
    \vec{\mathcal{R}}_{HB}=\begin{bmatrix}
        \mathcal{R}_{HB_1} \\
        \mathcal{R}_{HB_2} \\
        \vdots \\
        \mathcal{R}_{HB_n}
    \end{bmatrix} &= \left(\frac{1}{\gamma_{H}+\mu_{H}}\right)\boldsymbol{\beta_{B}}\boldsymbol{\Lambda_{HB}}\frac{\vec{B}^{*}}{B^{*}}.
\end{align}
With these definitions, we see that $\mathcal{R}_0 = \sqrt{\vec{\mathcal{R}}_{HB}\cdot\vec{\mathcal{R}}_{BH}}$. This expression in terms of type reproduction numbers can be used to better understand the overall contribution to infection dynamics of particular biting stages \citep{Heesterbeek2007-ty}.

\section{Case study: Comparing models of mosquito biting}\label{sec:case_study}
Here, we exhibit how different models, represented by various formulations of the mosquito biting process, can produce inconsistent predictions of how changes to mosquito biting affect the basic reproduction number, \Rzero{}. 
All calculations were conducted in \textit{Julia} or \textit{R} and all figures generated with the \textit{R} ggplot package \citep{R-Core-Team2025-kk, Wickham2016-or, Bezanson2017-nd}. 
All code used to generate data and figures in this article is available in the GitHub repository at https://github.com/kydahl/mosquito-bite-process-modeling.

\subsection{Model types}
To assess the impact on the basic reproduction number, \Rzero{}, we consider three formulations of the mosquito biting process from three different lenses: standard, empirical, and phenomenological. 
In particular, we fix the average GCD across the models to facilitate direct comparison. 
The standard exponential model of mosquito biting exhibits the property that \Rzero{} is inversely correlated to average GCD, that is, $\mathcal{R}_0 \propto 1/\textrm{GCD}$ \citep{Tedrow2019-gy}. 
We show numerically that this property is not universal among models of the mosquito biting process. 
This analysis is supplemented with a sensitivity analysis of \Rzero{} to the parameters introduced in the mechanistic model. 
From a control perspective, these suggest that, in order to understand the effect of perturbations of mosquito biting on transmission potential, one must be careful to select a model for the biting process that incorporates the relevant mechanisms affected by the perturbations.

The models will differ in the construction of the matrix $\boldsymbol{A}$ and vector $\vec{\alpha}$. 
Unless stated otherwise, all models use the parameter values specified in Table \ref{tab:parameters}. 
Recall that the GCD for a given representation of the biting process is given by equation \eqref{eq:GCD}. 
Because the latter two terms in equation \eqref{eq:GCD} are independent of the parameterization, we use the first term, $\theta = \vec{\alpha}^{T}(-\boldsymbol{A})^{-1}\vec{1}$, the mean of the distribution $\operatorname{PH}(\vec{\alpha}, \boldsymbol{A})$, to make comparisons across the models. 
For all models, we assume that the contact rate to the hosts is proportional to the ratio of biting vectors to hosts. 
In addition, the total host and mosquito populations are at their stable equilibrium with $S_H+I_H+R_H=K_H$ and $\vec{1}^T\vec{B} = B^*$.

\paragraph{Standard model}
To compare commonly used models for mosquito-borne disease transmission, we consider a model with no explicit oviposition or resting stages and an exponentially distributed waiting time for the biting process. 
The biting rate parameter, $b$, is assumed to be equal to the inverse of the GCD. 
For completeness, we write the full model here
\begin{equation}\label{eq:standard_model}
\begin{alignedat}{2}
&\frac{d}{dt}S_{H}&&=\mu_{H}K_{H}-\varPhi_{BH}S_{H}-\mu_{H}S_{H} ,\\
&\frac{d}{dt}I_{H}&&=\varPhi_{BH}S_{H}-\gamma_{H}I_{H}-\mu_{H}I_{H},\\
&\frac{d}{dt}R_{H}&&=\gamma_{H}I_{H}-\mu_{H}R_{H},\\
&\frac{d}{dt}J&&=\varphi\left(V,J\right)-\left(\rho_{J}+\mu_{J}\right)J,\\
&\frac{d}{dt}S&&=\rho_{J}J-\varPhi_{HB} S-\mu S,\\
&\frac{d}{dt}E&&=\varPhi_{HB} S -\eta E-\mu E,\\
&\frac{d}{dt}I&&= \eta E - \mu I.
\end{alignedat}
\end{equation}
In this case, the contact rates are $\Lambda_{BH} = bB/H$ and $\Lambda_{HB} = b$, and the transmission probabilities are \mbox{$\beta_H = \beta_B = 8.75\times10^{-2}$}. 
These transmission probability values were chosen to scale the \Rzero{} values from this model to facilitate visual comparison with the others. 
The forces of infection are then
$\varPhi_{BH} = b\beta_H\left(\frac{B^*}{K_H}\right)\frac{I}{B}$ and $\varPhi_{HB} = b\beta_B\frac{I_H}{K_H}$. 
We note that system \eqref{eq:standard_model} is equivalent to the generalized system \eqref{eq:epi_model} in the limit as $\gamma_R\to\infty$ and $\gamma_V\to\infty$ and $\boldsymbol{A} = -b$ and $\vec{\alpha} = 1$. 
Hence, $\Gamma_I=1/\mu$, $\Gamma_E = 1/(\mu+\eta)$, leading to $\tau_E=\eta/(\mu+\eta)$
and \mbox{$\Rzero=b\sqrt{\left(\frac{1}{\mu_H+\gamma_H}\right)\beta_H \frac{1}{\mu}\left(\frac{\eta}{\mu+\eta}\right) \beta_B \frac{B^*}{K_H}}$}, and the common finding that \Rzero{} is inversely proportional to the gonotrophic cycle duration.

\paragraph{Exponential model}
We consider a one-dimensional version of system \eqref{eq:epi_model}. 
Let $\boldsymbol{A} = -c$ and $\vec{\alpha}=1$. 
The contact rates are given by $\Lambda_{BH} = cB^*/K_H$, $\Lambda_{HB} = c$. In this case, $\theta = 1/c$. 
While the sole difference between this and the previous model is the inclusion of oviposition and resting states, the expression for \Rzero{} is much more complicated in this case because we must account for infectiousness in the mosquito developing while in the ovipositing or resting states. 

\paragraph{Empirical model}
Without actual data, we consider a scenario using simulated data to represent actual experiments. 
Suppose that data were collected on $\theta_s$, the biting state duration of mosquitoes in a laboratory setting. 
Suppose further that it is found that the heterogeneity among the individual mosquitoes of $\theta_s$ is best fit by a log-normal distribution, $\operatorname{LogNormal}(\mu_{\textrm{log}},\sigma_{\textrm{log}}^2)$. 
We chose to use a log-normal distribution to exemplify the flexibility of our framework because it is a positive distribution that cannot be directly expressed as a phase-type distribution, unlike Erlang or Coxian distributions. 

With this toy scenario, we consider what happens as the average biting state duration, $\theta$, corresponding to the mean of the log-normal distribution, is varied. 
For simplicity, we assume that the variance in the log-normal distribution is fixed at 1, i.e., $\sigma_{\textrm{log}}^2 = \ln(1+1/\theta^2)$. 
Then $\mu_{\textrm{log}} = \ln(\theta) - \sigma_{\textrm{log}}^2 / 2$. 
For each $\theta$ value, one hundred random samples are drawn from these distributions to be used as virtual data. 
A phase-type distribution of dimension $n=8$ is then fit to these data using the standard E-M algorithm provided in the \texttt{mapfit} package in \textit{R} \citep{Okamura2015-le, Bladt2021-pc}. 

With the $\vec{\alpha}$ and $\boldsymbol{A}$ fit to the synthetic data as described above, the contact rates are determined from the exit rate vector: \mbox{$\boldsymbol{\Lambda_{BH}} = \operatorname{diag}(-\boldsymbol{A}\vec{1})B^*/K_H$} and $\boldsymbol{\Lambda_{HB}} = \operatorname{diag}(-\boldsymbol{A}\vec{1})$. 
The probabilities are given by \mbox{$\boldsymbol{\beta_H} = \beta_H \boldsymbol{I}_8$} and \mbox{$\boldsymbol{\beta_B} = \beta_B\boldsymbol{I}_8$}.

\paragraph{Phenomenological model}
Inspired by \citet{Scott2000-wg}, we consider a scenario with individual heterogeneity among mosquitoes in terms of the number of bites they make within a single gonotrophic cycle. 
Using the results for Thailand in Table 1 of \cite{Scott2000-wg}, we assume that 53\% of all mosquitoes take one bite per gonotrophic cycle, 42\% take two bites, and 5\% take three bites. 
We assume that all types of mosquitoes have the same average biting stage duration, $\theta = 1/b$.  
We apply the linear chain trick to obtain transient rate matrices for each class of mosquito, then take their direct sum to obtain $\alpha_F=[0.53, 0.42, 0, 0.05,0,0]^T$ and $\boldsymbol{A_F}$ as defined in equation \eqref{eq:fate_PH}.
It can be verified that $\theta = \vec{\alpha}^T\left(-\boldsymbol{A_F}\right)^{-1}\vec{1} = \frac{1}{b}$. 

The contact rates are given here by \mbox{$\boldsymbol{\Lambda_{BH}} = -\operatorname{Diag}(\boldsymbol{A})B^*/K_H$} and \mbox{$\boldsymbol{\Lambda_{HB}} = -\operatorname{Diag}(\boldsymbol{A})$}, where $\operatorname{Diag}(\boldsymbol{M})$ indicates the column vector composed of the diagonal entries of $\boldsymbol{M}$. The transmission probabilities are given by $\boldsymbol{\beta_H} = \beta_H \boldsymbol{I}_6$ and $\boldsymbol{\beta_B} = \beta_B\boldsymbol{I}_6$.

\begin{figure}[t!]
    \centering
    \includegraphics[width=0.75\textwidth, trim={0  1.4in 3.45in 0},clip]{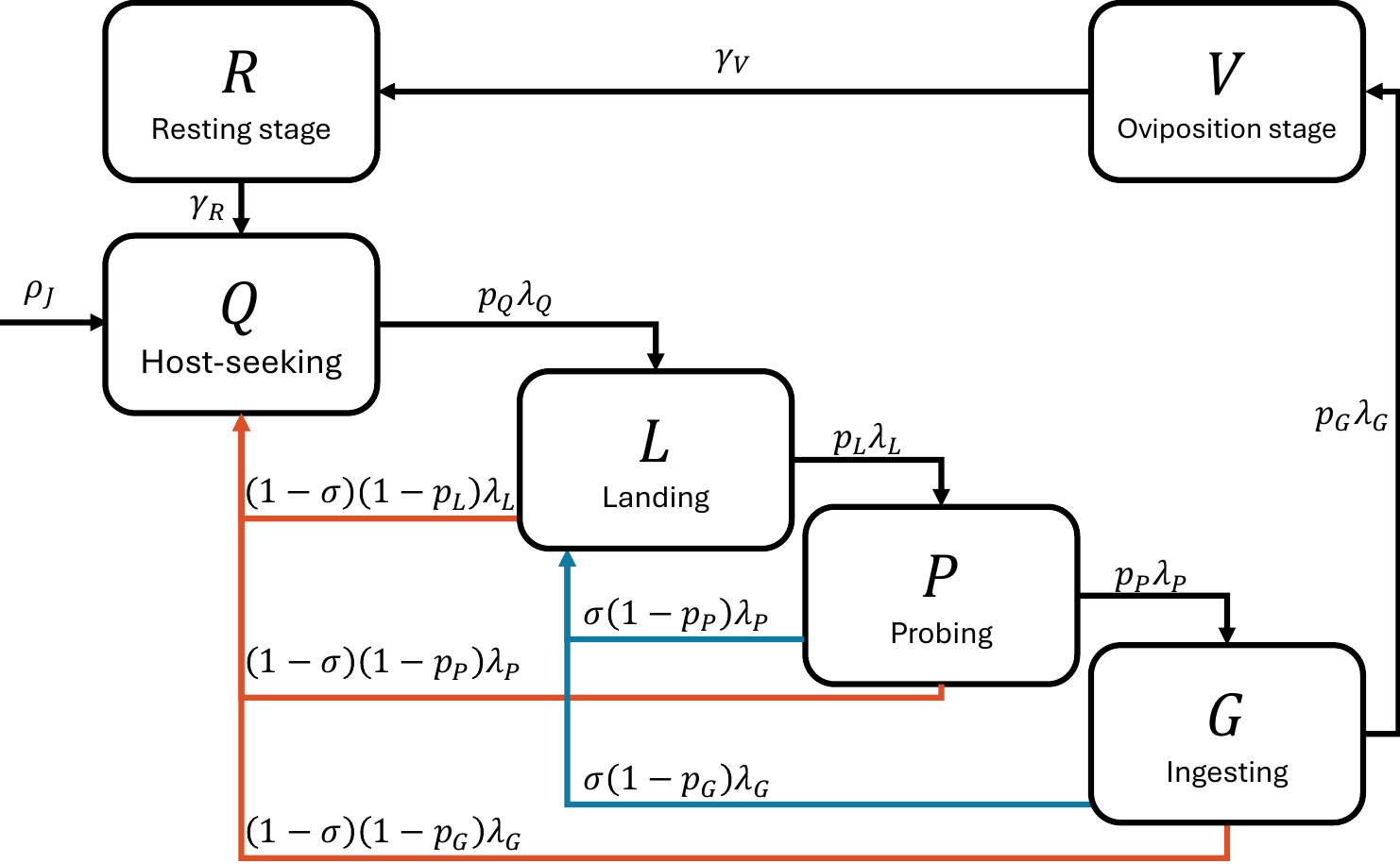}
    \caption{Compartmental diagram for the mechanistic model described in Section \ref{subsubsec:mech_model}. Newly emerged mosquitoes initially enter the host-seeking state. If they are successful throughout the biting process, mosquitoes proceed from host-seeking ($Q$) to landing on the host ($L$), then probing for a blood vessel ($P$), and finally ingesting blood ($G$), before leaving the host to find a location to lay eggs ($V$). Mosquitoes may be unsuccessful in their biting attempts, leading them to reattempt landing on the same host (blue lines) or seek out a new host (red lines). For visual clarity, the juvenile compartment, mortality rates, and self-loops are not drawn.}
    \label{fig:mech_pop_diagram}
\end{figure}

\paragraph{Mechanistic model}
\label{subsubsec:mech_model}
Finally, we consider a model that explicitly follows the mechanistic process described in Section \ref{sec:biting_process}. 
We sub-divide the population of biting mosquitoes, $B$, into four sub-compartments: host-seeking ($Q$); landing ($L$); probing ($P$); and ingesting ($G$). 
Successful progression out of state $X$ is determined by a single rate $\lambda_X$ and a single probability of success $p_X$. 
When mosquitoes do not successfully progress out of a state, they may return to their current or previous state. 
Mosquitoes disrupted during the landing or probing stages have a probability, $\sigma$, which we call the persistence probability, of attempting to re-land on the same host rather than seeking a new host. 
Figure \ref{fig:mech_pop_diagram} is a compartmental diagram of the process described by this model.

Letting $\vec{B}=\left[Q, L, P, G \right]^T$, this model is represented by
$\vec{\alpha}=\left[1, 0, 0, 0
\right]^T$,  and
\[
\boldsymbol{A}=\begin{bmatrix}-\lambda_{Q}+\left(1-p_{Q}\right)\lambda_{Q} & p_Q \lambda_Q & 0 & 0\\
\left(1-\sigma\right)\left(1-p_{L}\right)\lambda_{L} & -\lambda_{L}+\sigma\left(1-p_{L}\right)\lambda_{L} & p_{L}\lambda_{L} & 0\\
\left(1-\sigma\right)\left(1-p_{P}\right)\lambda_{P} & \sigma\left(1-p_{P}\right)\lambda_{P} & -\lambda_{P} & p_{p}\lambda_{P}\\
\left(1-\sigma\right)\left(1-p_{G}\right)\lambda_{G} & \sigma\left(1-p_{G}\right)\lambda_{G} & 0 & -\lambda_{G}
\end{bmatrix}.
\]
The definitions for the parameters for this model are given in Table \ref{tab:mech_parameters}. 
We considered two parameter sets with the same transition rates but varying transition probabilities. 
The first describes a hypothetical mosquito species, labeled `Flighty,' which has a high tendency to depart from a host when disturbed. 
The second hypothetical species is `Persistent' and is less likely to be disrupted from feeding and, if it is, tends to persist in its feeding attempts on the same host. 

For this model, instead of fitting $\boldsymbol{A}$ to match a given value of $\theta$, we vary each biting parameter independently to elicit changes in $\theta$. 
In particular, we will vary the host-seeking rate ($\lambda_Q$), probing success probability ($p_P$), and ingestion success probability ($p_G$). 

\begin{table}[t!]
    \centering    
    \caption{Parameters for the mechanistic model described in section \ref{subsubsec:mech_model}. The `Flighty' parameter set represents a hypothetical mosquito species that is easily disrupted and highly likely to restart the biting process and seek a new host. In contrast, the `Persistent' mosquito is less likely to be disrupted, and if it is, it is more likely to attempt to re-land on the same host.} 
    \label{tab:mech_parameters}
    \begin{tabular}{>{\centering}m{0.085\textwidth}m{0.56\textwidth}m{0.085\textwidth}m{0.085\textwidth}}
\hline
Parameter & Description & Flighty & Persistent \tabularnewline
\hline
$p_Q$ & Seeking success  & 100\% & 100\% \tabularnewline
& Probability of progressing from host-seeking $Q$ to landing $L$ & \tabularnewline
$\lambda_Q$ & Seeking rate& $1/(8 \text{ hr})$ & $1/(8 \text{ hr})$ \tabularnewline
& Exit rate from host-seeking stage $Q$  & \tabularnewline
$p_L$ & Landing success & 50\%  & 70\% \tabularnewline
& Probability of progressing from landing $L$ to probing $P$ & \tabularnewline
$\lambda_L$ & Landing rate & $1/(10 \text{ min})$   & $1/(10 \text{ min})$  \tabularnewline
&  Exit rate from landing stage $L$ & \tabularnewline
$p_P$ & Probing success & 50\%  & 80\% \tabularnewline
& Probability of progressing from probing $P$ to ingesting $G$  & \tabularnewline
$\lambda_P$ & Probing rate & $1/(5 \text{ min})$  & $1/(5 \text{ min})$ \tabularnewline
& Exit rate from probing stage $P$ & \tabularnewline
$p_G$ & Ingesting success & 50\%  & 90\% \tabularnewline
&  Probability of progressing from ingesting $G$ to ovipositing $V$ & \tabularnewline
$\lambda_G$ & Ingesting rate $G$ & $1/(1 \text{ min})$  & $1/(1 \text{ min})$ \tabularnewline
& Exit rate from ingestion stage & \tabularnewline
$\sigma$ & Persistence probability  & 10\%  & 34\% \tabularnewline
& Probability of attempting to re-land on the same host after disruption & \tabularnewline
\hline
    \end{tabular}
\end{table}

We make several assumptions regarding contact and infection that align with the mechanistic approach. 
These assumptions are drawn from the mosquito biting process described in Section \ref{sec:biting_process}. 
First, vertebrate hosts can only become infected if contacted by an infectious probing mosquito: $\beta_{H,P}=\beta_H$ and \mbox{$\beta_{H,Q}=\beta_{H,L}=\beta_{H,G}=0$}. 
Mosquitoes can only become exposed if they feed on an infectious vertebrate host: $\beta_{B,G}=\beta_B$ and $\beta_{B,Q}=\beta_{B,L}=\beta_{B,P}=0$. 
For illustration purposes, when ingestion success probability ($p_G$) is varied, we set \mbox{$\beta_H = \beta_B = 8.75\times10^{-2}$} to ensure that \Rzero{} is at a similar scale to the other models. 
Infectious contact only occurs when a mosquito exits a stage where the relevant transmission probability is positive: $\lambda_{H,P}(B^*,K_H) \propto \lambda_P$ and $\lambda_{B,G}(B^*,K_H) \propto \lambda_G$. 
From these assumptions, we arrive at the transmission probability and contact rate matrices,
\begin{align*}
    \boldsymbol{\beta_H} =     \begin{bmatrix}
        0 & 0 & 0 & 0\\
        0 & 0 & 0 & 0\\
        0 & 0 & \beta_{H} & 0\\
        0 & 0 & 0 & 0
    \end{bmatrix}\textrm{, }\boldsymbol{\beta_B} = \begin{bmatrix}
            0 & 0 & 0 & 0\\
            0 & 0 & 0 & 0\\
            0 & 0 & 0 & 0\\
            0 & 0 & 0 & \beta_{B}
        \end{bmatrix}\textrm{, }
    \boldsymbol{\Lambda_{BH}} =  \begin{bmatrix}
        0 & 0 & 0 & 0\\
        0 & 0 & 0 & 0\\
        0 & 0 & \lambda_{P}\frac{B^*}{K_H} & 0\\
        0 & 0 & 0 & 0
    \end{bmatrix}\textrm{, and } \boldsymbol{\Lambda_{HB}} = \begin{bmatrix}
            0 & 0 & 0 & 0\\
            0 & 0 & 0 & 0\\
            0 & 0 & 0 & 0\\
            0 & 0 & 0 & \lambda_{G}
        \end{bmatrix}.
\end{align*}
Under these assumptions, the forces of infection for the mechanistic model can be fully written out as:
\begin{equation}\label{eq:mech_forces_of_infection}
    \begin{aligned}
    \varPhi_{BH}	S_H&=\beta_{H}\lambda_{P}\frac{B^*}{K_H}\frac{I_{P}}{B}S_H,\\
\boldsymbol{\varPhi_{HB}}\vec{S}_B&=\begin{bmatrix}0\\
    0\\
    0\\
    \beta_{B}\lambda_{G}S_{G}
    \end{bmatrix}\frac{I_{H}}{K_H}.
\end{aligned}
\end{equation}
Finally, the basic reproduction number for this model instance can be expressed in terms of the type reproduction numbers as $\mathcal{R}_0 = \sqrt{\mathcal{R}_{PH}\mathcal{R}_{HG}}$, where \mbox{$\mathcal{R}_{PH} =\beta_{H}\lambda_{P}\frac{B^*}{K_{H}}\left[\boldsymbol{\Gamma_{I}}^T\boldsymbol{\tau_{E}}^T\right]_{3,4}$} is the type reproduction number from probing mosquitoes to hosts and $\mathcal{R}_{HG}=\left(\frac{1}{\gamma_{H}+\mu_{H}}\right)\beta_{B}\lambda_{G}\frac{G^{*}}{B^*}$ is the type reproduction number from hosts to ingesting mosquitoes. 

\subsection{Relating $\mathcal{R}_0$ and the gonotrophic cycle duration}
\begin{figure}[t!]
    \centering
    \includegraphics[width=\textwidth]{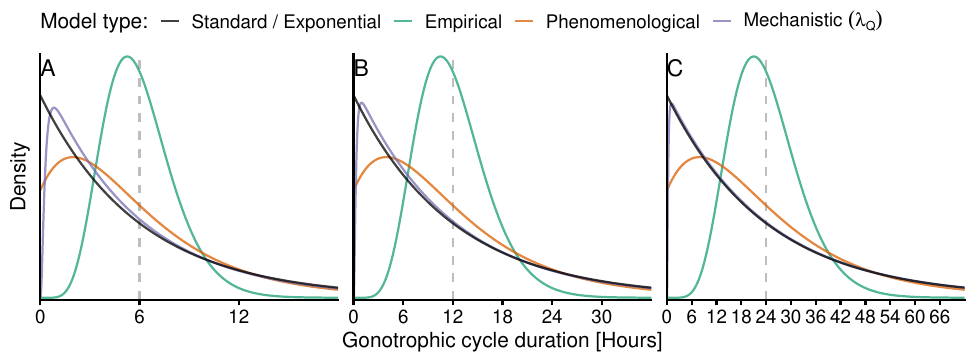}
    \caption{
    Probability density functions (PDFs) for the gonotrophic cycle duration for the five model types. 
    The distributions are compared by setting them to have equal means ($\theta$, see equation \eqref{eq:theta_dfn}, represented by a dashed gray line): six hours (A), twelve hours (B), and twenty-four hours (C). 
    See section \ref{subsubsec:mech_model} for details on how the mechanistic model probability density function is approximated for the given mean. 
    The PDFs associated with the standard and exponential models are equivalent. 
    The distributions for the mechanistic model obtained from changing $\lambda_Q$, $p_P$, and $p_G$ were visually identical, so only the curve for $\lambda_Q$ is shown here. 
    For the mechanistic model, baseline values for `Persistent' type mosquitoes were used (see Table \ref{tab:mech_parameters}). 
    All other parameters are given in Table \ref{tab:parameters}. 
    }
    \label{fig:compare_dists}
\end{figure}

The models' associated dwell-time distributions can appear markedly distinct, as viewed through their probability density functions (PDF), even though the model parameterization is adjusted to have the same mean GCD (Figure \ref{fig:compare_dists}). 
Surprisingly, the PDFs of the mechanistic model obtained from varying different parameters largely overlapped (a single purple curve is shown in Fig. \ref{fig:compare_dists}).
The PDFs of the standard and exponential models are identical and represented by a black curve. 
However, the overlap in PDFs is not as significant for very short average gonotrophic cycle durations (Figure \ref{fig:compare_dists}A). 
The empirical model has a PDF that closely approximates a log-normal distribution (orange curve), which is to be expected since it was initially fitted to a log-normal distribution. 
The empirical model PDF also exhibits the greatest difference in shape from the others. 
Assuming a single bite per gonotrophic cycle, a common modeling assumption, the PDFs indicate that the standard, exponential, and mechanistic models should exhibit very similar population and transmission dynamics. 
In contrast, the empirical and phenomenological models display distinct dynamics. 

\begin{figure}[t!]
    \centering
    \includegraphics[width=\textwidth, trim={0 0 0 0},clip]{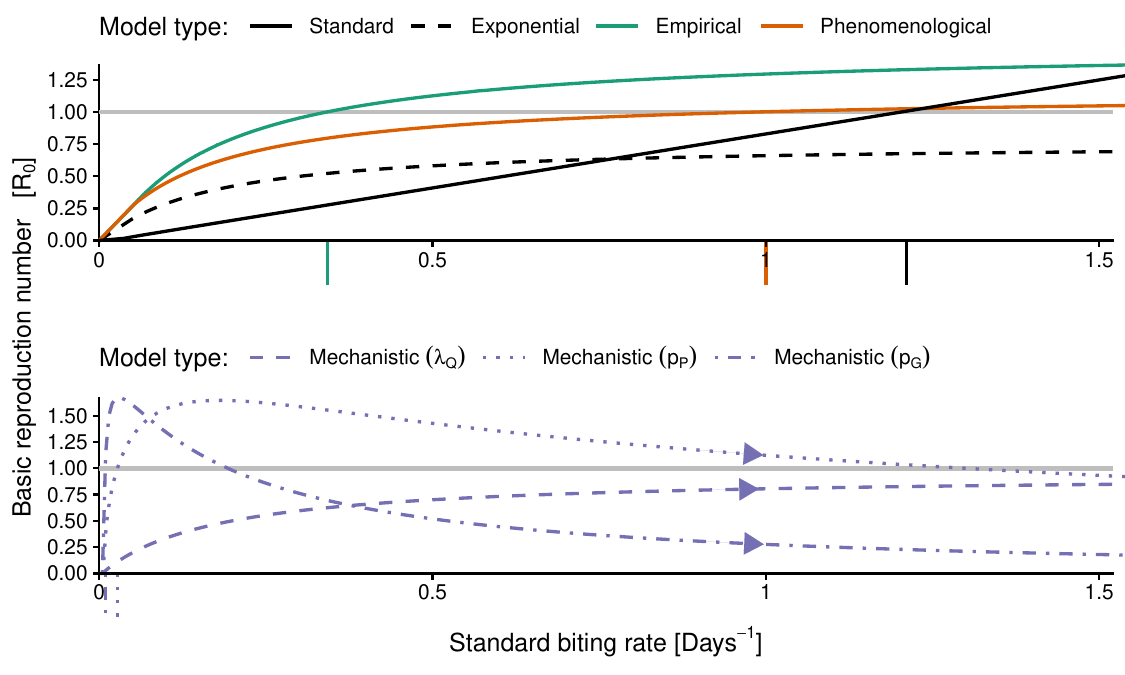}
    \caption{Relationships between \Rzero{} and the standard biting rate, the inverse of the gonotrophic cycle duration, for each model type. 
    Tick marks indicate the critical minimum standard biting rate, the standard biting rate at which \Rzero{} first exceeds one (see Table \ref{tab:R0_summary}). 
    The gray horizontal line indicates where \Rzero{} equals one. 
    For the mechanistic models, the standard biting rate increases as the parameters listed in parentheses are increased (indicated by arrowheads along the curves). 
    The mechanistic model uses parameters from the `Persistent' mosquito parameter set in Table \ref{tab:mech_parameters}. 
    The $\beta_H$ and $\beta_B$ values for the standard and the mechanistic ($p_G$) models were reduced to 8.75\% to highlight the qualitative differences among the model types better visually. 
    All other parameters are given in Table \ref{tab:parameters}. 
    }
    \label{fig:invGCD_R0}
\end{figure}
\Rzero{} determines the short and long-term dynamics of system \eqref{eq:epi_model}, regardless of the specific parameterization of the biting process (Theorem \ref{thm:DFE_stability}). 
We, therefore, use \Rzero{} to directly compare the transmission dynamics of the various biting process models. 
In particular, we examine the relationship between \Rzero{} and the inverse of the average biting stage duration, $\theta$, which we refer to as the `standard biting rate', a quantity typically assumed to be the mosquito biting rate in modeling, though this assumption is often implicit \citep{Zahid2023-ne}. 

Matching expectations, the standard model \Rzero{} exhibits a linear relationship to the standard biting rate (black line, Figure \ref{fig:invGCD_R0}). 
In the other models, the relationship between \Rzero{} and the standard biting rate takes one of two forms: unimodal with a unique maximum (as in the mechanistic models varying $p_P$ or $p_G$, purple dotted and dot-dash curves, respectively, in Figure \ref{fig:invGCD_R0}) or a monotonically increasing saturating function. 
Arrows on the \Rzero{} curves of the mechanistic model indicate that the standard biting rate increases as the parameters are increased (dashed, dotted, and dot-dashed curves, Figure \ref{fig:invGCD_R0}). 
Notably, although the PDF of the exponential model closely resembled that of the standard model, the shape of its \Rzero{} curve is notably different from that of the standard model (dashed black and solid black curves, respectively, Figure \ref{fig:invGCD_R0}).

The critical minimum standard biting rate, the value at which \Rzero{} first exceeds one and disease persists, varies substantially across models. 
For the standard model, \Rzero{} exceeds one when the standard biting rate is greater than approximately 1.2 bites per day (Table \ref{tab:R0_summary}).
For the exponential model and the mechanistic model varying $\lambda_Q$, \Rzero{} does not exceed one for realistic parameter values and remains below one even up to a standard biting rate of 20 bites per day. 
While the critical minimum standard biting rate for the mechanistic ($p_P$) and ($p_G$) models is substantially lower than the others ($0.0273$ and $0.00972$, respectively), these curves also exhibit critical \textit{maximum} standard biting rates, of 1.30 and 0.19, respectively, above which \Rzero{} falls below one. 
Decreases in the GCD may occur alongside even greater decreases in the multiple biting number, leading to an overall reduction in transmission.

The curves for \Rzero{} of the exponential, empirical, phenomenological, and mechanistic ($\lambda_Q$) models approach horizontal asymptotes ($0.759$, $1.52$, $1.15$, and $0.939$, respectively) because the ovipositing and resting stages limit how quickly mosquitoes can complete the cycle. 
Unlike the standard model, for these models, no matter how quickly the biting process proceeds, a mosquito always spends an average of seven days finding an oviposition site, ovipositing, then resting and finding nutritional resources to support another transit through the biting process. 

\begin{table}[t!]
    \centering    
    \caption{Summary characteristics of the \Rzero{} curves in Figure \ref{fig:invGCD_R0}. 
    Values are calculated for standard biting rate values ranging from zero to twenty bites per day. 
    Maximum \Rzero{} values indicate the largest value of \Rzero{} attained as the standard biting rate was varied from zero to 1.5 bites per day. 
    The critical minimum standard biting rate is the smallest value of the standard biting rate at which \Rzero{} exceeds one. 
    The critical maximum standard biting rate is the largest value of the standard biting rate at which \Rzero{} is less than one after initially exceeding it. 
    Dashes indicate that the value does not apply to the given model type. 
    } 
    \label{tab:R0_summary}
    \begin{tabular}{lccc}
\hline
& & \multicolumn{2}{l}{Standard biting rate} \tabularnewline
Model type & Maximum \Rzero{} & Critical minimum & Critical maximum\tabularnewline
\hline
    Standard                  & $60.1$  & $1.21$   & ---     \tabularnewline
    Exponential               & $0.759$ & ---      & ---     \tabularnewline
    Empirical                 & $1.52$  & $0.342$  & ---     \tabularnewline
    Phenomenological          & $1.15$  & $1$      & ---     \tabularnewline
    Mechanistic ($\lambda_Q$) & $0.939$ & ---      & ---     \tabularnewline
    Mechanistic ($p_P$)       & $1.65$  & $0.0491$ & $1.30$  \tabularnewline
    Mechanistic ($p_G$)       & $1.61$  & $0.0484$ & $0.190$ \tabularnewline
    \hline
    \end{tabular}
\end{table}

The non-monotonic relationships exhibited by two of the mechanistic models indicate that \Rzero{} is not always increased by reductions in GCD. 
Lower success probabilities ($p_P$ or $p_G$) lead to longer GCDs and, thus, lower standard biting rates. 
At the same time, more feeding failures necessitate more frequent biting attempts, increasing contact rates between mosquitoes and hosts and, therefore, \Rzero{}. 
This indicates a trade-off between GCD and multiple biting that contrasts with the independence of GCD and multiple biting implicitly assumed in most models \citep{Tedrow2019-gy,Zahid2023-ne}. 

\subsection{Sensitivity analysis of the mechanistic model}
\begin{figure}[t!]
    \centering
    \includegraphics[width=\textwidth, trim={0 0.2in 0.1in 0},clip]{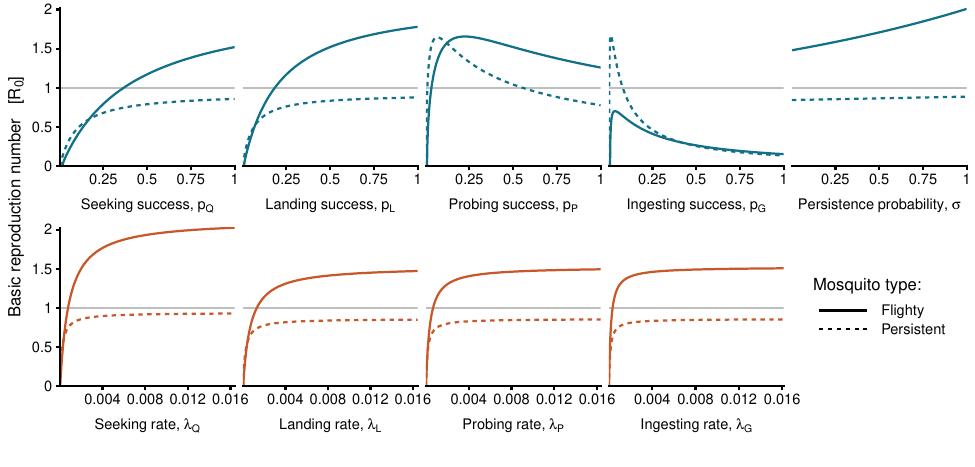}
    \caption{
    The relationships between \Rzero{} and the parameters of the mechanistic model described in section \ref{subsubsec:mech_model}, with parameters representing hypothetical `Flighty' (solid curves) and `Persistent' (dashed curves) mosquito species. 
    The labeled parameter is varied for each subplot while all others are held fixed to their values given in Tables \ref{tab:parameters} and \ref{tab:mech_parameters}. 
    Probabilities vary from zero to one, and rates vary from zero to one per hour. 
    The gray horizontal line indicates where \Rzero{} equals one.
    For all rates, \Rzero{} has a horizontal asymptote. 
    }
    \label{fig:mech_vars_R0}
\end{figure}

While \Rzero{} can be expressed as a function of the mechanistic parameters with equation \eqref{eq:R0}, the analytical expression is too unwieldy to provide helpful information or warrant direct examination. 
To better understand the relationship between \Rzero{} and the mechanistic biting parameters, we evaluated \Rzero{} as each parameter was varied independently (Figure \ref{fig:mech_vars_R0}). 
There is inconsistency in how \Rzero{} responds to variations in the probability parameters; most increase monotonically, while the curves for $p_P$ and $p_G$ attain an initial peak before decreasing. 
However, for the rate parameters, the response of \Rzero{} is essentially the same: they all increase towards a horizontal asymptote.

Under most parameter combinations, the `Flighty' mosquito has higher \Rzero{} values than the `Persistent' mosquito. 
The exceptions occur at low values of probing success or ingesting success. 
The greater `Flighty' mosquito \Rzero{} indicates that higher levels of feeding disruption increase transmission, likely through an increase in multiple biting. 
While not included in this model, in nature, feeding disruption due to host defensiveness may lead to mosquito mortality and, therefore, some reduction in transmission. 
Notably, the \Rzero{} of the `Persistent' mosquito is very insensitive to the persistence probability ($\sigma$). 
This may be because persistence probability is only relevant when feeding failure occurs in the biting process, which is less likely for the `Persistent' mosquito.

\begin{figure}[t!]
    \centering
    \includegraphics[width=\textwidth,trim={0.175in 0.2in 0 0},clip]{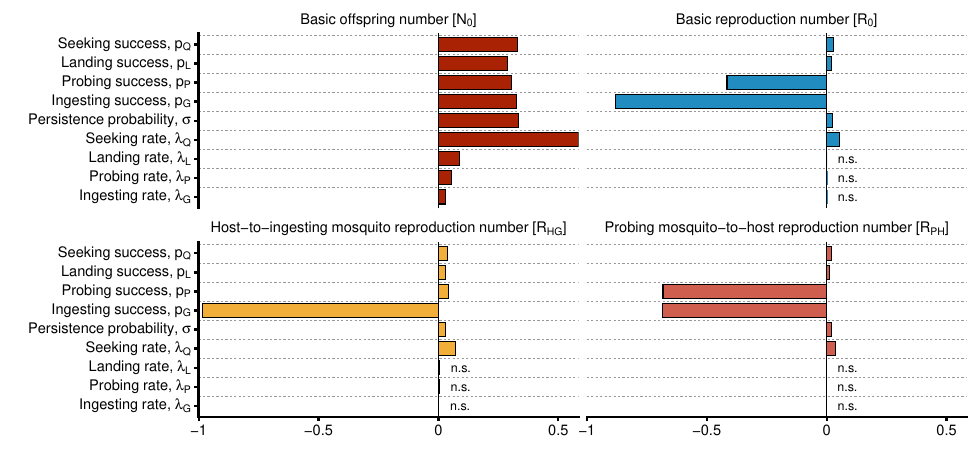}
    \caption{
    Partial rank correlation coefficients of the basic offspring number (red bars), basic reproduction number (blue bars), host-to-vector reproduction number (yellow bars), and vector-to-host reproduction number (pink bars) with respect to the mechanistic model biting parameters under a LHS-PRCC analysis. 
    To avoid situations where the basic reproduction number was undefined (that is, when the basic offspring number is less than one), probabilities were sampled uniformly from 20\% to 100\% except $\sigma$, which was varied from 1\% to 100\%. All rates were sampled uniformly from once every thirty minutes to twice per minute, except for the questing rate, which varied from once every eight hours to once every nine minutes.
    The label ``n.s.'' indicates sensitivity indices which were not significant when compared to a dummy parameter (as in \citet{Marino2008-av}).
    }
    \label{fig:all_PRCCs}
\end{figure}

To further explore the impacts of our parameter choices in the mechanistic model, we conducted a global sensitivity analysis of the basic offspring number, $\mathcal{N}_0$, basic reproduction number, \Rzero{}, host-to-ingesting mosquito reproduction number, $\mathcal{R}_{HG}$, and probing mosquito-to-host reproduction number, $\mathcal{R}_{PH}$. 
The mechanistic model parameters (Table \ref{tab:mech_parameters}) were sampled uniformly from the largest range of biologically reasonable values for which, additionally, $\mathcal{R}_0$ was defined (that is, when $\mathcal{N}_0>1$). All success probabilities were sampled uniformly between 20\% and 100\%, except for persistence probability, which was sampled uniformly between 1\% and 100\%. Rates were sampled between once every thirty minutes and once per thirty seconds, except for the host-seeking rate, which was sampled uniformly between once per three days and once per nine minutes. 
Latin hypercube sampling was performed within these ranges, resulting in a total of 100,000 samples.
We then calculated partial rank correlation coefficients between these parameters and $\mathcal{N}_0$, \Rzero{}, and the host-to-vector and vector-to-host type reproduction numbers (Figure \ref{fig:all_PRCCs}). 
We also conducted an eFAST global sensitivity analysis \citep{Marino2008-av,Wu2013-tj} with the same parameters and outputs, which produced results similar to the PRCC analysis (see Appendix \ref{GSA:eFAST}). 
All sensitivity analysis calculations were conducted in \textit{Julia} \citep{Bezanson2017-nd}.

The basic offspring number, $\mathcal{N}_0$, was positively influenced by all the mechanistic model parameters. 
As defined, these parameters are all associated with more rapid or more successful traversal through the biting stages.
$\mathcal{N}_0$ was most sensitive to the host-seeking rate ($\lambda_Q$), indicating the importance of the ability to find and locate hosts for the persistence of mosquito populations. 
The success probability parameters had a roughly equal positive influence on $\mathcal{N}_0$.

The basic reproduction number was most strongly impacted by the probing and ingesting success probabilities, indicating that when mosquitoes are more likely to fail in the probing or ingesting stages, $\mathcal{R}_0$ is larger. 
This reflects that if a mosquito is disrupted while probing or ingesting, it must return once again to the probing and ingesting stages, leading to more opportunities for transmission. 
A closer look at the type reproduction numbers indicates that ingesting success has an extremely strong negative influence on the host-to-ingesting mosquito reproduction number and a mildly negative effect on the probing mosquito-to-host reproduction number. 
The probing and ingesting success probabilities have a roughly equal negative influence on the probing mosquito-to-host reproduction number. 
When a mosquito fails at ingesting, it must return again to the probing stage in order to eventually attain a complete blood meal. 
Because of this, failing at probing or failing at ingesting has a roughly equivalent effect of allowing for another transmission event.  
All of the rates except the host-seeking rate had an insignificant effect on the reproduction numbers.

There is an inconsistency between the parameters that best support mosquito population persistence and those that increase disease transmission. 
In particular, transmission is increased when mosquitoes fail at ingesting or probing (necessitating additional blood meals to transition to ovipositing), while population persistence, measured through $\mathcal{N}_0$, is increased when mosquitoes are more successful in transiting these stages. 
These tradeoffs are reflected in the local sensitivity analysis (Figure \ref{fig:mech_vars_R0}), where $\mathcal{R}_0$ peaks at low values of $p_P$ and $p_G$ and rapidly approaches zero as they are further decreased, likely due to the effect of these parameters on $\mathcal{N}_0$. 


\section{Discussion}
The mosquito biting process is the driving force of vector-borne disease transmission. 
In contrast to the incidental transmission of environmentally and directly transmitted pathogens, mosquito-borne pathogens commandeer the mosquitoes' adept skill at finding and feeding on vertebrate hosts to reproduce. 
Empiricists, therefore, seek improved methods for measuring mosquito biting rates, and modelers develop innovative transmission models to understand how biting rate drives transmission. 
Most commonly, an implicit assumption is that a single bite occurs in a single gonotrophic cycle, which is incorporated as a constant biting rate \citep{Zahid2023-ne}. 
Still, other models of the biting process are often built to address specific questions or systems, limiting the generality of findings and their ability to compare results across models. 
We developed a generalized framework that can be used to model an extensive range of possible forms of the mosquito biting process.

The framework can incorporate representations of various behavioral phenomena driving mosquito contact rates, including multiple biting and host defensive behaviors. 
The framework accommodates any perspective taken to address modeling goals, whether empirical, phenomenological, or mechanistic. 
The forces of infection must be specified separately from the model of the biting process, but in many cases, they can be derived from knowledge of the biting process. 
Tractable formulas for the basic offspring number and basic reproduction number allow for analysis of how individual-level parameters translate into population-level, long-term disease outcomes. 
The average duration spent in the biting stages, $\theta$, can be used to facilitate comparisons among different models and to elucidate how assumptions about the biting process (independent of the time spent biting) lead to varying levels of transmission potential, as measured by \Rzero{}.

Considering several models as a case study, we exhibited the flexibility of the framework and its ability to facilitate comparisons among different representations of the biting process. 
In particular, different biting process models led to markedly different relationships between \Rzero{} and the biting stage duration. 
In contrast to the linear relationship derived from the standard model, more complex biting process models can lead to increasing, saturating, or even non-monotonic relationships. 
This case study showed that it is incorrect to assume that decreasing the gonotrophic cycle duration necessarily increases transmission potential. 

Multiple biting and the GCD are commonly considered independent quantities in models. 
However, as the results for our mechanistic model indicated, there can be important trade-offs between these two quantities. 
If mosquitoes take a fixed number of bites per gonotrophic cycle, a shorter GCD indicates higher contact rates and, therefore, higher transmission levels. 
However, if the number of bites per gonotrophic cycle is allowed to vary (e.g., due to disruption to the biting process), then there is a trade-off between GCD and the number of bites per gonotrophic cycle. 
Higher failure rates lead to higher contact rates as mosquitoes bite more frequently to attain a sufficient blood meal. 
While these trade-offs may be plausible, they lack experimental or observational evidence. 
Modern imaging techniques allow for measuring mosquito response to disturbances while feeding \citep{Wynne2020-cl,Wynne2022-cp}. 
Thus, experimental studies could establish how mosquito biting rates are altered as the rate of disturbance increases. 

The ability to incorporate a strong relationship between multiple biting and GCD in models has particular relevance for studies of the effect of rising temperatures on mosquito-borne pathogen transmission. 
Shifts in temperature lead to concomitant changes in mosquito life history traits and, therefore, the rate at which they proceed through their gonotrophic cycle \citep{Mordecai2019-md}. 
More needs to be uncovered about how aspects of the biting process change due to shifts in temperature. 
For example, how does temperature affect a mosquito's ability to locate and home in on a host? 
Are mosquitoes more or less easily disrupted when temperatures rise? 
Complicating things further, the answers to these questions are likely species-dependent. 
Future studies should aim to determine key biting-related rates and probabilities as temperature is varied in a laboratory setting. 

Our framework provides a tool for studying the effects of feeding disruption and host defensive behaviors on overall transmission. 
In particular, it provides the groundwork for investigating the effects of mosquito feeding failure on population-level transmission dynamics, as demonstrated in both mechanistic and phenomenological models. 
In nature, feeding failure can be due to host defensive behaviors that are effective at killing biting mosquitoes \citep{Edman1972-wp}. 
This trade-off between fecundity and mortality may be an essential driver of the evolved differences in mosquito biting behavior \citep{Klowden1994-su}. 
A crucial next step is to utilize this modeling framework to assess the impact of increased mortality rates resulting from feeding failure on overall transmission. 

As the emergence rate of zoonoses increases \citep{Jones2008-uv}, identifying the actual wildlife reservoirs of disease is critical for preventing or suppressing outbreaks \citep{Haydon2002-cc}. 
A mechanistic representation of the effects of host defensive behaviors on the mosquito biting process could elucidate which vertebrate hosts are more likely to be the reservoirs of mosquito-borne disease. 
Small rodents are often implicated as reservoirs of mosquito-borne viruses such as Venezuelan Equine Encephalitis virus and La Crosse virus \citep{Harding2018-ck, Deardorff2009-vu}.
However, observational and experimental evidence indicate that small rodents are extremely effective defenders against mosquito biting, often through predation on the attacking mosquitoes \citep{Day1984-ph}. 
These small rodents may be susceptible to infection by mosquito-borne viruses, but their contribution to overall transmission is poorly understood. 
Our modeling framework provides a unique platform for scaling up individual-level empirical measurements of the defensive behaviors of small rodents and their effectiveness in disrupting mosquito biting to population-level predictions of transmission potential. 
This may be used to generate hypotheses on whether populations of small rodents can feasibly be the cause of mosquito-borne virus outbreaks or whether they are incidental carriers of these viruses. 

Mechanistic models of the biting process can also provide vital information about the effectiveness of personal protection measures for reducing mosquito-borne pathogen transmission. 
While the effectiveness of repellents at the individual level can be determined straightforwardly, ascertaining their effect at the population level is non-trivial. 
Even if an individual is more protected, it does not mean that overall transmission is necessarily diminished.
For example, repellents may simply push mosquitoes off to less-protected populations and, therefore, have no overall impact on population-level transmission. 
By explicitly modeling the effect of repellent on particular aspects of mosquito biting, our models can be used to estimate their impact on reducing \Rzero{}. 
This could also provide recommendations for improving repellents. 
For example, our sensitivity analysis results suggest that repellents that repel mosquitoes only after they have begun probing could actually lead to a large increase in overall transmission. 
Such an increase in transmission would be difficult to overcome by either decreasing the landing success probability or the host-seeking rate, which both have a relatively much smaller effect on decreasing transmission.

The mosquito biting process is a key component of mosquito-borne disease transmission. 
Here, we provide a generalized framework for incorporating the mosquito biting process into transmission models in a straightforward and tractable manner. 
Such a framework has the potential to enhance the realism of mosquito-borne disease models, allowing for more precise incorporation of biting behavior and more biologically accurate insight. 
Alongside new technological developments that allow for fine-scale measurements of mosquito behavior in the laboratory, modelers and empiricists are poised to greatly expand our understanding of the relationship between mosquito biting behaviors and the emergence and persistence of disease outbreaks.

\backmatter

\section*{Acknowledgements}
The authors thank Clément Vinauger for insightful discussions and valuable feedback on the description of mosquito biology in this manuscript, and two anonymous reviewers for their thorough and helpful comments. This material is based upon work supported by the National Science Foundation MPS-Ascend Postdoctoral Research Fellowship under Grant No. MPS-2316455 to KJMD. MAR was supported in part by a Burroughs Wellcome Fund Climate and Health Interdisciplinary Award (BWF \#1022621). LMC was supported in part by National Science Foundation Grant DMS-2144680.

\section*{Conflict of interest statement}
We declare no conflict of interest associated with this publication.

\section*{Author Contributions}
Conceptualization: Kyle J.-M. Dahlin, Lauren M. Childs, Michael A. Robert; Methodology: Kyle J.-M. Dahlin, Lauren M. Childs, Michael A. Robert; Formal analysis and investigation: Kyle J.-M. Dahlin; Writing - original draft preparation: Kyle J.-M. Dahlin; Writing - review and editing: Kyle J.-M. Dahlin, Lauren M. Childs, Michael A. Robert; Funding acquisition: Kyle J.-M. Dahlin, Lauren M. Childs; Resources: Lauren M. Childs; Supervision: Lauren M. Childs.

\section*{Data Availability Statement}
All data and code necessary to replicate the results in this article are available at \\https://github.com/kydahl/mosquito-bite-process-modeling

\bibliography{MBF}
\clearpage


\begin{appendices}

\renewcommand{\thesubsection}{\Alph{subsection}}
\section*{Appendices}

\renewcommand\thefigure{S\arabic{figure}}  

\subsection{Proof of Theorem 1}\label{proof:positive_equilibria}
To determine the existence of a positive equilibrium of system \eqref{eq:orig_model}, we solve the system of equations \eqref{eq:equilibria} under the assumption that $V$, $J$, or at least one of the entries of $\vec{B}$ is non-zero. 
\begin{align}\label{eq:equilibria}
0 & =\varphi\left(V^*,J^*\right)-\left(\rho_{J}+\mu_{J}\right)J^*\nonumber,\\
0 & =\left(\rho_{J}J^*+\gamma_{R}R^*\right)\vec{\alpha}+\boldsymbol{A}^{T}\vec{B}^*-\mu\vec{B}^*\nonumber,\\
0 & =\left(-\boldsymbol{A}\vec{1}\right)^{T}\vec{B}^*-\gamma_{V}V^*-\mu V^*,\nonumber\\
0 &=\gamma_{V}V^*-\gamma_{R}R^*-\mu R^*.
\end{align}

Let $r=\vec{B}^{T}\left(-\boldsymbol{A}\vec{1}\right)$, noting that $r\ge0\in\mathbb{R}$), and $\tau=\vec{\alpha}^T\left(\mu\boldsymbol{I}-\boldsymbol{A}\right)^{-1}\left(-\boldsymbol{A}\vec{1}\right)$.
Now since $V^{*}=\frac{1}{\gamma+\mu}r$ and $R^{*}=\left(\frac{1}{\mu+\gamma_{R}}\right)\frac{\gamma_{V}}{\mu+\gamma_{V}}r$, we obtain from the second equation above
\[J^{*}=\frac{1}{\tau\rho_{J}}\left[1-\tau\left(\frac{\gamma_{R}}{\mu+\gamma_{R}}\right)\left(\frac{\gamma_{V}}{\mu+\gamma_{V}}\right)\right]r,\]
and then, from the first equation along with assumption J2,
\newcommand{\eqdef}{=\mathrel{\mathop:}}
\begin{equation}\label{eq:eq_exist}
    \bar{\varphi}\left(J^{*}\right)=\left[\tau\left(\frac{1}{\mu+\gamma_{V}}\right)\left(\frac{\rho_{J}}{\rho_{J}+\mu_{J}}\right) n_{G}\right]^{-1}
\end{equation}
where $\bar{\varphi}\left(J^{*}\right)= {\varphi}\left(V^*,J^{*}\right) / V^*$, $\rho=\tau\frac{\gamma_{R}}{\mu+\gamma_{R}}\frac{\gamma_{V}}{\mu+\gamma_{V}}$ and $n_{G}=\frac{1}{1-\rho}$. 
By assumption J3, there is at least one solution, $J^*$, to equation \eqref{eq:eq_exist}. 
Given $J^*$, the other equilibrium values can be determined as follows
\begin{align*}
    \vec{B}^{*}	&= \left(\mu\boldsymbol{I}-\boldsymbol{A}^{T}\right)^{-1}\vec{\alpha}\left(1+\left(\frac{\gamma_{R}}{\mu+\gamma_{R}}\right)\left(\frac{\gamma_{V}}{\mu+\gamma_{V}}\right)\tau n_{G}\right)\rho_{J}J^{*},\\
    V^{*} &= \left(\frac{1}{\mu+\gamma_{V}}\right)\tau n_{G}\rho_{J}J^{*},\\
    R^{*} &= \tau n_{G}\left(\frac{1}{\mu+\gamma_{R}}\right)\left(\frac{\gamma_{V}}{\mu+\gamma_{V}}\right)\rho_{J}J^{*}.
\end{align*}
Reversing the process, we see that any non-trivial solution of \eqref{eq:equilibria}, $(J^*, \vec{B}^*, V^*,R^*)$, must also satisfy equation \eqref{eq:eq_exist}.

\subsection{Considerations for specifying the transmission model}\label{sec:specific_transmission_model}
\subsubsection{Transmission probability}\label{sec:transmission_probability}
Infectious mosquitoes may not be capable of transmitting pathogens at all stages of biting, and similarly, susceptible mosquitoes may not become infected at all biting stages. 
For example, transmission to vertebrate hosts may only be possible when infectious mosquitoes are probing for blood vessels. 
In contrast, transmission to mosquitoes can occur when they begin feeding on the blood of an infectious host \citep{Thongsripong2021-hv}. 
We, therefore, decompose the biting stages $B_1$, $\ldots$, $ B_n$ into subsets. 
As in the previous subsection, let $\Omega\subseteq \{B_1,\ldots, B_n\}$ be the set of biting stages where contact occurs, with $\Omega_H \subseteq \Omega$ the stages where transmission to the host can occur, and $\Omega_B \subseteq \Omega$ the stages where transmission to the mosquito can occur. 
Define $\beta_{H,i}$ as the probability that contact with an infectious mosquito in biting stage $i$ leads to infection in a susceptible vertebrate host. 
Then $\beta_{H,i}>0$ if and only if $B_i\in\Omega_H$. 
Define the diagonal matrix for host susceptibility to be $\boldsymbol{\beta_H} = \operatorname{diag}([\beta_{H,1},\ldots,\beta_{H,n}])$. 
Similarly, we define $\beta_{B,j}$ as the probability that a susceptible mosquito in biting stage $j$ becomes infected after contact with an infectious host, where $\beta_{B,j}>0$ if and only if $B_j\in\Omega_B$. 
Then the matrix for mosquito susceptibility is $\boldsymbol{\beta_B} = \operatorname{diag}([\beta_{B,1},\ldots,\beta_{B,n}])$. 
Note that the quantities $\beta_{H,i}$ or $\beta_{B,j}$ can depend on both the infector's infectivity and the infected's susceptibility. 
\subsubsection{Contact rates}\label{sec:contact_rates}
For simplicity, we only consider types of contact that can result in transmission (this restriction is also enforced by the transmissibility term later). 
We define $\boldsymbol{\Lambda_{BH}}$ to be the host contact rate matrix that has the property that $\vec{1}^T\boldsymbol{\Lambda_{BH}}\vec{B}$ gives the total rate at which vertebrate hosts are contacted by all biting state mosquitoes. 
We define $\boldsymbol{\Lambda_{HB}}$ to be the mosquito contact matrix such that $\boldsymbol{\Lambda_{HB}}\vec{B}$ is the vector of the rates at which biting state mosquitoes make contact with hosts.  
For now, we write $\boldsymbol{\Lambda}$ to refer to these matrices generally and return to the distinction between $\boldsymbol{\Lambda_{BH}}$ and $\boldsymbol{\Lambda_{HB}}$ in the following subsection. 

To decide how to formulate these rates, one must consider three questions: which biting stages are associated with host contact?; at what point of the contact process can transmission occur?; does host availability limit the contact rate?.

Not all biting stages contribute to the contact rate between mosquitoes and vertebrate hosts. 
For example, a biting stage might represent intermediate steps in the biting process, like the time a mosquito spends seeking a host or the time spent resting between feeding attempts. 
We define the set $\Omega$ to contain all stages where contact can occur. 
Then, if stage $B_i$ is not associated with contact, $B_i\not\in\Omega$ and each element of the $i^{\textrm{th}}$ row of $\boldsymbol{\Lambda}$ is set to zero. 

Transmission may occur at different points of the contact process. 
For some infectious agents (e.g., West Nile virus), transmission can occur at the onset of contact \citep{Styer2007-br}, while, for others, successful transmission might occur only when the mosquito has become engorged \citep{Blanken2024-yn}. 
We consider three cases in which contact is associated with a biting stage $B_i$: (i) contact occurs while the mosquito is in stage $B_i$; (ii) contact occurs when the mosquito enters $B_i$; (iii) contact occurs when the mosquito exits $B_i$. 

Case i is the most general and straightforward. 
One must decide on the rate, $c_i$, at which mosquitoes in $B_i$ make contact with vertebrate hosts, which need not depend on the rates in the transient rate matrix $A$. 
Let $\boldsymbol{\Lambda} = \textrm{diag}\left(c_i\right)$ where $c_i>0$ only if $B_i\in\Omega$.

In case ii, the entrance rate into $B_j$ from $B_i$ is obtained from the transient rate matrix \boldA{} as $a_{ij}B_i$. 
The matrix of entrance rates to $B_j$ is therefore given by the diagonal matrix whose entries are the elements of column $j$ of \boldA{} with a zero in row $j$, that is, 
$\boldsymbol{\Lambda}_j = \operatorname{diag}([c_{1}, \ldots, c_{n}])$ where $c_{i} = a_{ij}$ if $i\neq j$ and zero otherwise. 
Combining these rates across all contact states $B_j\in \Omega$, we obtain $\boldsymbol{\Lambda} = \sum_{B_j\in\Omega} \boldsymbol{\Lambda}_j$. 

In case iii, infectious contact for mosquitoes is initiated only when exiting a biting stage $B_i$ where the relevant rate is $-a_{ii}B_i$.
The full contact rate matrix is then \mbox{$\boldsymbol{\Lambda}= \operatorname{diag}([c1,\ldots,cn])$} where $c_i = -a_{ii}$ if $B_i\in\Omega$ and $c_i = 0$ otherwise.  
Alternatively, if we assume that any mosquito leaving the biting state must have made infectious contact with a host, then we would set  $\boldsymbol{\Lambda}=\textrm{diag}(-\boldsymbol{A}\vec{1})$. 

It is commonly assumed that the availability of hosts---their abundance and tolerance of mosquito biting---limits contact rates. 
Vertebrate hosts commonly exhibit intolerance to the feeding or presence of biting insects, manifesting in defensive behaviors that limit the rate at which contact is made \citep{Edman1988-am}. 
A similar effect can also occur if only a limited amount of surface area on hosts allows for feeding by mosquitoes so that mosquitoes interfere with each other's ability to feed successfully \citep{Kershenbaum2012-qr}. 
Various methods for incorporating host availability into contact rates have been explored \citep{Alto2001-nq,Bowman2005-kj, Chitnis2006-kn, Blayneh2010-ah, Thongsripong2021-hv}. 

A critical step for using any of these methods is ensuring that ``conservation of biting'' is maintained. 
At each point in time, the number of bites that the vertebrate host population encounters is equal to the number of bites made by the mosquito population \citep{Blackwood2018-cy}. 
In general, the per-capita contact rate for mosquitoes in stage $i$, $\lambda_{B,i} \left(\vec{B},H\right)$, and for vertebrate hosts, $\lambda_H \left(\vec{B},H\right)$, can take on different values as long as the balance equation 
\begin{equation}\label{eq:bite_balance}
\lambda_H \left(\vec{B},H\right) H = \sum_{i=1}^n\lambda_{B,i} \left(\vec{B},H\right) B_i
\end{equation}
is satisfied for the scenario where mosquitoes feed only on the focal vertebrate host \citep{Blackwood2018-cy}. 

One common assumption in mosquito-borne disease models is that the rate at which an individual host is contacted by a vector is reduced when either the vector population is very low or the host population is very large. 
This is usually incorporated in models by having the per-capita contact rate experienced by vertebrate hosts be proportional to the vector-host ratio, while, in contrast, the per-capita contact rate for mosquitoes is constant and independent of both mosquito and host densities. 
In this case, this common assumption would be represented by setting $\lambda_{B,i} \left(B,H\right) = \lambda_{B,i}$ and $\lambda_H \left(\vec{B},H\right) = \sum_{i=1}^n \lambda_{B,i}B_i/H$, which can also be obtained by dividing both sides of equation \eqref{eq:bite_balance} by $H$. 
After deciding which stages are associated with contact, whether transmission occurs with entry, exit, or residence in those stages, and whether host availability limits the contact rate, we ultimately arrive at the two contact rate matrices, $\boldsymbol{\Lambda_{BH}} = \left[\lambda_{H,{ij}}\left(\vec{B},H\right)\right]_{1\le i,j\le n}$ and \mbox{$\boldsymbol{\Lambda_{HB}} =\left[\lambda_{B,{ij}}\left(\vec{B},H\right)\right]_{1\le i,j\le n}$}.

\subsubsection{Probability of contact with an infectious individual}\label{sec:prob_contact_w_infective}
We assume that infectious individuals are well-mixed in their respective populations so that the probability of selecting an infectious individual out of that population is equal to the proportion of infectious individuals. 
Therefore, the probability that a host has been contacted by a mosquito in biting stage $i$ is $B_i/B$ and the probability that that mosquito is infectious is ${I_i}/B_i$, so that the total probability of a host being contacted by an infectious mosquito in biting stage $i$ is $I_i/B$. 
Similarly, $I_H/H$ is the probability that the host with which a mosquito is making contact is infectious.

\subsection{Interpretation of the probability $\tau$}\label{proof:interp_tau}
We wish to provide further intuition to interested readers in the interpretation of \mbox{$\tau=\vec{\alpha}^{T}\left[\left(\mu\boldsymbol{I}-\boldsymbol{A}\right)^{-1}\left(-\boldsymbol{A}\vec{1}\right)\right]$} as the total probability of surviving and exiting the biting stages. 
We start by verifying that $\tau$ is bounded between zero and one. 

By construction of the matrix $\boldsymbol{A}$, the matrix $\left(\mu\boldsymbol{I}-\boldsymbol{A}\right)^{-1}$ and the vector $-\boldsymbol{A}\vec{1}$ are both positive elementwise. 
Since $\vec\alpha$ is also positive elementwise, $\tau\geq0$. 
Now, when $\mu=0$, $\tau=1$ since $\vec\alpha^T\vec1=1$. 
Finally, $\tau$ is monotonically decreasing in $\mu$ because
\begin{equation*}
\frac{d\tau}{d\mu} = -\vec{\alpha}^{T}\left[\left(\mu\boldsymbol{I}-\boldsymbol{A}\right)^{-2}\left(-\boldsymbol{A}\vec{1}\right)\right]
\end{equation*}
is always negative since the vector and matrix are both positive elementwise, by the same arguments as above. 

In the basic case where $A=-b$ and $\alpha = 1$, we see that $\tau=\frac{b}{\mu+b}$. 
This is the ratio of the ``successful'' exit rate from the biting stage, $b$, to the total rate out of the biting stage, $\mu+b$. 
Equivalently, this is the product of the ``successful'' exit rate and the average duration spent in the biting stage, $1/\left(\mu+b\right)$. Generalizing to two dimensions, consider the case where there are two types of mosquitoes that bite at the rates $b_1$ and $b_2$, and mosquitoes are assigned these types with probabilities $\alpha_1$ and $\alpha_2$, respectively. In this case,
\begin{equation*}
\boldsymbol{A} =  \begin{bmatrix}
            -b_1 & 0 \\
            0 & -b_2
        \end{bmatrix}\textrm{, and }\vec\alpha^T=\left[\alpha_1,\alpha_2\right].
\end{equation*} 
Then \mbox{$\tau = \alpha_{1}\left(\frac{b_{1}}{\mu+b_{1}}\right)+\alpha_{2}\left(\frac{b_{2}}{\mu+b_{2}}\right)$} is an average of the two survival probabilities weighted by the probabilities of belonging to the two types.

The general case can be understood in analogy to these simple cases. 
The vector $-\boldsymbol{A}\vec{1}$ provides the ``successful'' exit rates out of the biting stages. 
The entries, $c_{ij}$, of $\left(\mu\boldsymbol{I}-\boldsymbol{A}\right)^{-1}$, give the average duration an individual that started in stage $i$ will eventually spend in stage $j$. 
Thus, the $i$th entry of the vector \mbox{$\left(\mu\boldsymbol{I}-\boldsymbol{A}\right)^{-1}\left(-\boldsymbol{A}\vec{1}\right)$} is the total probability of exiting the biting state successfully through biting stage $i$.  
Finally, $\vec{\alpha}^T$ gives the ``weights'' to this average, according to the probabilities of starting in the different stages.

\subsection{Proof of Theorem 2}\label{proof:stability_extinction}
We can calculate the basic offspring number of the population in
this general form and determine the stability of the extinction equilibrium. 
We use the next-generation matrix method described in \citep{van-den-Driessche2002-ck}. 

Define a new individual as any entering the biting stages (i.e. developmentally reproductively mature) for the first time. Then, we obtain the next-generation matrix from the decomposition
\begin{align*}
\frac{d}{dt}\begin{bmatrix}J\\
\vec{B}\\
V\\
R
\end{bmatrix}&=\begin{bmatrix}0\\
\rho_{J}J\vec{\alpha}\\
0\\
0
\end{bmatrix}-\left\{ \begin{bmatrix}\left(\mu_{J}+\rho_{J}\right)J\\
-\boldsymbol{D}_{\textrm{diag}\boldsymbol{A}}^{T}\vec{B}+\mu\vec{B}\\
\left(\mu+\gamma_{V}\right)V\\
\left(\mu+\gamma_{R}\right)R
\end{bmatrix}-\begin{bmatrix}\varphi\left(V,J\right)\\
\gamma_{R}R\vec{\alpha}+\left(\boldsymbol{A}-\boldsymbol{D}_{\textrm{diag}A}\right)^{T}\vec{B}\\
\left(-\boldsymbol{A}\vec{1}\right)^{T}\vec{B}\\
\gamma_{V}V
\end{bmatrix}\right\}  \\\frac{d}{dt}\begin{bmatrix}J\\
\vec{B}\\
V\\
R
\end{bmatrix}&=\mathcal{F}-\mathcal{V}
\end{align*}
Let $F$ be the Jacobian of $\mathcal{F}$ evaluated at the extinction equilibrium $\left(0,\vec{0},0,0\right)$ and let $V$ the Jacobian of $\mathcal{V}$ evaluated at the extinction equilibrium. Then the next-generation matrix is $\boldsymbol{K}=\boldsymbol{F}\boldsymbol{V}^{-1}$ where $\boldsymbol{F}$ and $\boldsymbol{V}$ are given below.
\begin{align*}
\boldsymbol{F}&=\begin{bmatrix}0 & \boldsymbol{0}_{1\times n} & 0 & 0\\
\rho_{J}\vec{\alpha} & \boldsymbol{0}_{n\times n} & \boldsymbol{0}_{n\times1} & \boldsymbol{0}_{n\times1}\\
0 & \boldsymbol{0}_{1\times n} & 0 & 0\\
0 & \boldsymbol{0}_{1\times n} & 0 & 0
\end{bmatrix},
\\\boldsymbol{V}&=\begin{bmatrix}\left(\rho_{J}+\mu_{J}\right)-\phi_{J} & \boldsymbol{0}_{1\times n} & -\phi_{V} & 0\\
\boldsymbol{0}_{n\times1} & \mu\boldsymbol{I}-\boldsymbol{A}^{T} & \boldsymbol{0}_{n\times1} & -\gamma_{R}\vec{\alpha}\\
0 & \left(\boldsymbol{A}\vec{1}\right)^{T} & \left(\mu+\gamma_{V}\right) & 0\\
0 & \boldsymbol{0}_{1\times n} & -\gamma_{V} & \left(\mu+\gamma_{R}\right)
\end{bmatrix}
\end{align*}
where $\boldsymbol{0}_{m\times n}$ is a matrix of size $m$ by $n$ whose entries are all zero. 
Here we have defined $\phi_{J}=\frac{\partial\varphi}{\partial J}\left(0,0\right)$ and $\phi_{V}=\frac{\partial\varphi}{\partial V}\left(0,0\right)$. 

We first need to find the inverse of the matrix $\boldsymbol{V}$, which we will compute by decomposing $\boldsymbol{V}$ into block matrices. Let 
\begin{align*}
&\boldsymbol{W}=\begin{bmatrix}\left(\rho_{J}+\mu_{J}\right)-\phi_{J} & \boldsymbol{0}_{1\times n}\\
\boldsymbol{0}_{n\times1} & \mu\boldsymbol{I}-\boldsymbol{A}^{T}
\end{bmatrix},
&&\boldsymbol{X}=\begin{bmatrix}-\phi_{V} & 0\\
\boldsymbol{0}_{n\times1} & -\gamma_{R}\vec{\alpha}
\end{bmatrix},\\
&\boldsymbol{Y}=\begin{bmatrix}0 & -\left(-\boldsymbol{A}\vec{1}\right)^{T}\\
0 & \boldsymbol{0}_{1\times n}
\end{bmatrix}, &&\boldsymbol{Z}=\begin{bmatrix}\left(\mu+\gamma_{V}\right) & 0\\
-\gamma_{V} & \left(\mu+\gamma_{R}\right)
\end{bmatrix},
\end{align*}
so that $\boldsymbol{V}$ can be written in block matrix form as $\boldsymbol{V}=\begin{bmatrix}\boldsymbol{W} & \boldsymbol{X}\\
\boldsymbol{Y} & \boldsymbol{Z}
\end{bmatrix}$. 
Since $\boldsymbol{W}$ is invertible ($\left(\mu \boldsymbol{I}-\boldsymbol{A}^{T}\right)$ is the transient rate matrix for the minimum of two independent phase-type distributions), we obtain
\[\boldsymbol{V}^{-1}	=\begin{bmatrix}\boldsymbol{W}^{-1}+\boldsymbol{W}^{-1}\boldsymbol{X}\left(\boldsymbol{Z}-\boldsymbol{Y}\boldsymbol{W}^{-1}\boldsymbol{X}\right)^{-1}\boldsymbol{Y}\boldsymbol{W}^{-1} & -\boldsymbol{W}^{-1}\boldsymbol{X}\left(\boldsymbol{Z}-\boldsymbol{Y}\boldsymbol{W}^{-1}\boldsymbol{X}\right)^{-1}\\
-\left(\boldsymbol{Z}-\boldsymbol{Y}\boldsymbol{W}^{-1}\boldsymbol{X}\right)^{-1}\boldsymbol{Y}\boldsymbol{W}^{-1} & \left(\boldsymbol{Z}-\boldsymbol{Y}\boldsymbol{W}^{-1}\boldsymbol{X}\right)^{-1}
\end{bmatrix}.\]
Using the definitions of $\tau$ and $n_G$ from the Proof of Theorem 1, we can express $\left(\boldsymbol{Z}-\boldsymbol{Y}\boldsymbol{W}^{-1}\boldsymbol{X}\right)^{-1}$ as
\[
\left(\boldsymbol{Z}-\boldsymbol{Y}\boldsymbol{W}^{-1}\boldsymbol{X}\right)^{-1}=\begin{bmatrix}n_{G}\frac{1}{\mu+\gamma_{V}} & \tau n_{G}\left(\frac{\gamma_{R}}{\mu+\gamma_{R}}\right)\frac{1}{\mu+\gamma_{V}}\\
n_{G}\left(\frac{\gamma_{V}}{\mu+\gamma_{V}}\right)\frac{1}{\mu+\gamma_{R}} & n_{G}\frac{1}{\mu+\gamma_{R}}
\end{bmatrix}.
\]
Similarly computing the other entries of $\boldsymbol{V}^-1$, we eventually obtain
\[
\boldsymbol{V}^{-1}=\begin{bmatrix}\frac{1}{\left(\rho_{J}+\mu_{J}\right)-\phi_{J}} & \frac{\phi_{V}}{\left(\rho_{J}+\mu_{J}\right)-\phi_{J}}n_{G}\frac{1}{\mu+\gamma_{V}}\boldsymbol{M} & \frac{\phi_{V}}{\left(\rho_{J}+\mu_{J}\right)-\phi_{J}}n_{G}\frac{1}{\mu+\gamma_{V}} & 0\\
\boldsymbol{0}_{n\times1} & \left[\left(\mu\boldsymbol{I}-\boldsymbol{A}^{T}\right)^{-1}+n_{G}\left(\frac{\gamma_{V}}{\mu+\gamma_{V}}\right)\left(\frac{\gamma_{R}}{\mu+\gamma_{R}}\right)\boldsymbol{N}\boldsymbol{M}\right] & n_{G}\left(\frac{\gamma_{V}}{\mu+\gamma_{V}}\right)\left(\frac{\gamma_{R}}{\mu+\gamma_{R}}\right)\boldsymbol{N} & n_{G}\left(\frac{\gamma_{R}}{\mu+\gamma_{R}}\right)\boldsymbol{N}\\
0 & n_{G}\frac{1}{\mu+\gamma_{V}}\boldsymbol{M} & \frac{1}{\mu+\gamma_{V}} & \frac{1}{\mu+\gamma_{V}}\frac{\gamma_{R}}{\mu+\gamma_{R}}\tau\\
0 & n_{G}\left(\frac{\gamma_{V}}{\mu+\gamma_{V}}\right)\frac{1}{\mu+\gamma_{R}}\boldsymbol{M} & \frac{\gamma_{V}}{\mu+\gamma_{V}}\frac{1}{\mu+\gamma_{R}} & \frac{1}{\mu+\gamma_{R}}
\end{bmatrix}
\]
where, for the sake of space, we have made the substitutions \mbox{$\boldsymbol{M}=\left(-\boldsymbol{A}\vec{1}\right)^{T}\left(\mu\boldsymbol{I}-\boldsymbol{A}^{T}\right)^{-1}$} and \mbox{$\boldsymbol{N}=\left(\mu\boldsymbol{I}-\boldsymbol{A}^{T}\right)^{-1}\vec{\alpha}$}.

Now similarly decomposing $\boldsymbol{F}$ into matching block matrices, we obtain
\begin{align*}
\boldsymbol{F}=\begin{bmatrix}\boldsymbol{F}_{11} & \boldsymbol{0}_{n+1\times2}\\
\boldsymbol{0}_{2\times n+1} & \boldsymbol{0}_{2\times2}
\end{bmatrix},
\boldsymbol{F}_{11}=\begin{bmatrix}0 & \boldsymbol{0}_{1\times n}\\
\rho_{J}\vec{\alpha} & \boldsymbol{0}_{n\times n}
\end{bmatrix}
\end{align*}
Finally, calculating the next-generation matrix $\boldsymbol{K}=\boldsymbol{F}\boldsymbol{V}^-1$, we see that it is in fact upper block-triangular,
\[
\boldsymbol{K}	=\begin{bmatrix}\boldsymbol{F}_{11}\left[\boldsymbol{W}^{-1}+\boldsymbol{W}^{-1}\boldsymbol{X}\left(\boldsymbol{Z}-\boldsymbol{Y}\boldsymbol{W}^{-1}\boldsymbol{X}\right)^{-1}\boldsymbol{Y}\boldsymbol{W}^{-1}\right] & -\boldsymbol{F}_{11}\boldsymbol{W}^{-1}\boldsymbol{X}\left(\boldsymbol{Z}-\boldsymbol{Y}\boldsymbol{W}^{-1}\boldsymbol{X}\right)^{-1}\\
\boldsymbol{0}_{2\times n+1} & \boldsymbol{0}_{2\times2}
\end{bmatrix},
\]
so that its eigenvalues are zero (with multiplicity two) and the eigenvalues of \mbox{$\boldsymbol{Q}=\boldsymbol{F}_{11}\left[\boldsymbol{W}^{-1}+\boldsymbol{W}^{-1}\boldsymbol{X}\left(\boldsymbol{Z}-\boldsymbol{Y}\boldsymbol{W}^{-1}\boldsymbol{X}\right)^{-1}\boldsymbol{Y}\boldsymbol{W}^{-1}\right]$}. 
In fact, $\boldsymbol{Q}$ is itself lower block-triangular,
\[
\boldsymbol{Q}=\begin{bmatrix}0 & \boldsymbol{0}_{1\times n}\\
\frac{\rho_{J}}{\left(\rho_{J}+\mu_{J}\right)-\phi_{J}}\vec{\alpha} & \frac{\rho_{J}}{\left(\rho_{J}+\mu_{J}\right)-\phi_{J}}n_{G}\frac{\phi_{V}}{\mu+\gamma_{V}}\vec{\alpha}\left(-\boldsymbol{A}\vec{1}\right)^{T}\left(\mu\boldsymbol{I}-\boldsymbol{A}^{T}\right)^{-1}
\end{bmatrix},
\]
so its eigenvalues are zero and the eigenvalues of \mbox{$\boldsymbol{\tilde{Q}}=\frac{\rho_{J}}{\left(\rho_{J}+\mu_{J}\right)-\phi_{J}}n_{G}\frac{\phi_{V}}{\mu+\gamma_{V}}\vec{\alpha}\left(-\boldsymbol{A}\vec{1}\right)^{T}\left(\mu\boldsymbol{I}-\boldsymbol{A}^{T}\right)^{-1}$}. 
Finally, we apply the Weinstein-Aronszajn identity to find the eigenvalues of $\boldsymbol{\tilde{Q}}$,
\begin{align*}
\det\left(\lambda\boldsymbol{I}-\boldsymbol{\tilde{Q}}\right)&=\det\left(\lambda\boldsymbol{I}_{n}-\frac{\rho_{J}}{\left(\rho_{J}+\mu_{J}\right)-\phi_{J}}n_{G}\frac{\phi_{V}}{\mu+\gamma_{V}}\vec{\alpha}\left(-\boldsymbol{A}\vec{1}\right)^{T}\left(\mu\boldsymbol{I}-\boldsymbol{A}^{T}\right)^{-1}\right),\\\det\left(\lambda\boldsymbol{I}-\boldsymbol{\tilde{Q}}\right)&=\lambda^{n}\det\left(\boldsymbol{I}_{n}-\lambda^{-1}\frac{\rho_{J}}{\left(\rho_{J}+\mu_{J}\right)-\phi_{J}}n_{G}\frac{\phi_{V}}{\mu+\gamma_{V}}\vec{\alpha}\left(-\boldsymbol{A}\vec{1}\right)^{T}\left(\mu\boldsymbol{I}-\boldsymbol{A}^{T}\right)^{-1}\right),\\\det\left(\lambda\boldsymbol{I}-\boldsymbol{\tilde{Q}}\right)&=\lambda^{n}\det\left(1-\lambda^{-1}\frac{\rho_{J}}{\left(\rho_{J}+\mu_{J}\right)-\phi_{J}}n_{G}\frac{\phi_{V}}{\mu+\gamma_{V}}\left(-\boldsymbol{A}\vec{1}\right)^{T}\left(\mu\boldsymbol{I}-\boldsymbol{A}^{T}\right)^{-1}\vec{\alpha}\right),\\\det\left(\lambda\boldsymbol{I}-\boldsymbol{\tilde{Q}}\right)&=\lambda^{n-1}\left(\lambda-\frac{\rho_{J}}{\left(\rho_{J}+\mu_{J}\right)-\phi_{J}}\frac{\phi_{V}}{\mu+\gamma_{V}}\tau n_{G}\right).
\end{align*}
Here, we made the substitution $\tau = \vec{\alpha}^{T}\left[\left(\mu\boldsymbol{I}-\boldsymbol{A}\right)^{-1}\left(-\boldsymbol{A}\vec{1}\right)\right] = \left[\left(-\boldsymbol{A}\vec{1}\right)^{T}\left(\mu\boldsymbol{I}-\boldsymbol{A}^{T}\right)^{-1}\vec{\alpha}\right]^T$.

The eigenvalues of $\boldsymbol{K}$ are zero (multiplicity $n+2$) and \mbox{$\mathcal{N}_0 = \frac{\rho_{J}}{\left(\rho_{J}+\mu_{J}\right)-\phi_{J}}\frac{\phi_{V}}{\mu+\gamma_{V}}\tau n_{G}$}.
Thus, by Theorem 2 in \cite{van-den-Driessche2002-ck}, the extinction equilibrium is locally asymptotically stable if and only if $\mathcal{N}_0 <1$.

\subsection{Proof of Theorem 3}\label{proof:DFE_stability}
We show that the basic reproduction number $\mathcal{R}_0$ in \eqref{eq:R0} is the spectral radius of the next-generation matrix associated with system \eqref{eq:epi_model}. 

The full system of equations \eqref{eq:epi_model} has disease-free equilibria given by \mbox{$\left(S_{H},I_{H},R_{H},J,\vec{S}_{B},\vec{E}_{B},\vec{I}_{B},S_{V},E_{V},I_{V}S_{R},E_{R},I_{R}\right)=\left(K_{H},0,0,J^{\cdot},\vec{B}^{\cdot},\vec{0},\vec{0},V^{\cdot},0,0,R^{\cdot},0,0\right)$} for any $\left(J^{\cdot},\vec{B}^{\cdot},V^{\cdot},R^{\cdot}\right)$ that is an equilibrium of the base mosquito demographic model \eqref{eq:orig_model}. We assume that the base mosquito demographic model has a positive, non-extinction equilibrium, labeled $\left(J^{*},\vec{B}^{*},V^{*},R^{*}\right)$, and that this equilibrium is locally asymptotically stable in the absence of disease. Define $B^*=\vec{1}^{T}\vec{B}^{*}$. 
A disease-free equilibrium of \eqref{eq:epi_model} is given by $\left(I_{H},R_{H},\vec{E}_{B},\vec{I}_{B},E_{V},I_{V}E_{R},I_{R}\right)=\left(0,0,\vec{0},\vec{0},0,0,0,0\right)$.

Define a new infected individual to be any mosquito entering any of the exposed stages ($\vec{E}_B$, $E_V$, or $E_R$) or a host entering the infected stage ($I_H$). 
The system of equations \eqref{eq:epi_model} can now be decomposed as
\begin{align*}
    \frac{d}{dt}\begin{bmatrix}I_{H}\\
R_{H}\\
\vec{E}_{B}\\
\vec{I}_{B}\\
E_{V}\\
I_{V}\\
E_{R}\\
I_{R}
\end{bmatrix}	=\mathcal{F}-\mathcal{V}.
\end{align*}
where $\mathcal{F}$ is the new infection operator and $\mathcal{V}$ is the net transitions operator
\begin{align*}
\mathcal{F}&=\begin{bmatrix}\varPhi_{BH}\left(K_{H}-I_{H}-R_{H}\right)\\
0\\
\boldsymbol{\varPhi_{HB}}\left(\vec{B}^{*}-\vec{E}-\vec{I}\right)\\
\boldsymbol{0}_{n\times1}\\
0\\
0\\
0\\
0
\end{bmatrix},\\\mathcal{V}&=\begin{bmatrix}\left(\mu_{H}+\gamma_{H}\right)I_{H}\\
\mu_{H}R_{H}\\
\left(\mu+\eta\right)\vec{E}-\boldsymbol{D}_{\textrm{diag}\boldsymbol{A}}^{T}\vec{E}\\
\mu\vec{I}-\boldsymbol{D}_{\textrm{diag}\boldsymbol{A}}^{T}\vec{I}\\
\left(\mu+\gamma_{V}+\eta\right)E_{V}\\
\left(\mu+\gamma_{V}\right)I_{V}\\
\left(\mu+\gamma_{R}+\eta\right)E_{R}\\
\left(\mu+\gamma_{R}\right)I_{R}
\end{bmatrix}-\begin{bmatrix}0\\
\gamma_{H}I_{H}\\
\left(\gamma_{R}E_{R}\right)\vec{\alpha}+\left(\boldsymbol{A}-\boldsymbol{D}_{\textrm{diag}\boldsymbol{A}}\right)^{T}\vec{E}\\
\left(\gamma_{R}I_{R}\right)\vec{\alpha}+\eta\vec{E}+\left(\boldsymbol{A}-\boldsymbol{D}_{\textrm{diag}\boldsymbol{A}}\right)^{T}\vec{I}\\
\left(-\boldsymbol{A}\vec{1}\right)^{T}\vec{E}_{B}\\
\left(-\boldsymbol{A}\vec{1}\right)^{T}\vec{I}_{B}+\eta E_{V}\\
\gamma_{V}E_{V}\\
\eta E_{R}+\gamma_{V}I_{V}
\end{bmatrix}
\end{align*}
The symbol $\boldsymbol{0}_{m\times n}$ denotes the $m\times n$ matrix with entries all zero. 
Let $F$ be the Jacobian of $\mathcal{F}$ evaluated at the disease-free equilibrium and $V$ the Jacobian of $\mathcal{V}$ evaluated at the disease-free equilibrium. 

Our goal is to determine the spectral radius of $\boldsymbol{K}=\boldsymbol{F}\boldsymbol{V}^{-1}$.
As in the proof of Theorem 2, we will proceed by expressing $\boldsymbol{F}$ and $\boldsymbol{V}$ in block-matrix form, exhibiting that it takes on an upper- or lower- block-triangular form to reduce the problem of determining the eigenvalues of $\boldsymbol{K}$ to those of a lower-dimensional matrix. 
We will show that the eigenvalues of $\boldsymbol{K}$ are zeroes with some multiplicity and the eigenvalues of a $2\times2$ matrix $\boldsymbol{L}$.

The matrices $F$ and $V$ can be written in block-matrix form as
\[
\boldsymbol{F}=\begin{bmatrix}\boldsymbol{0}_{2\times2} & \boldsymbol{F_{VH}} & \boldsymbol{0}_{2\times4}\\
\boldsymbol{F_{HV}} & \boldsymbol{0}_{2n\times2n} & \boldsymbol{0}_{2n\times4}\\
\boldsymbol{0}_{4\times2} & \boldsymbol{0}_{4\times2n} & \boldsymbol{0}_{4\times4}
\end{bmatrix}, \textrm{ and }\boldsymbol{V}=\begin{bmatrix}\boldsymbol{V_{H}} & \boldsymbol{0}_{2\times2n+2} & \boldsymbol{0}_{2\times4}\\
\boldsymbol{0}_{2n+2\times2} & \boldsymbol{V_{BB}} & \boldsymbol{V_{RB}}\\
\boldsymbol{0}_{4\times2} & \boldsymbol{V_{BV}} & \boldsymbol{V_{VR}}
\end{bmatrix}
\]
where,
\[
\boldsymbol{F_{VH}}=\begin{bmatrix}\boldsymbol{0}_{1\times n} & \left(\frac{K_{H}}{B^{*}}\right)\vec{1}^{T}\boldsymbol{\beta_{H}}\boldsymbol{\Lambda_{BH}}\\
\boldsymbol{0}_{1\times n} & \boldsymbol{0}_{1\times n}
\end{bmatrix}\textrm{, }\boldsymbol{F_{HV}}=\begin{bmatrix}\frac{1}{K_{H}}\boldsymbol{\beta_{B}}\boldsymbol{\Lambda_{HB}}\vec{B}^{*} & \boldsymbol{0}_{n\times1}\\
\boldsymbol{0}_{n\times1} & \boldsymbol{0}_{n\times1}
\end{bmatrix},
\]
\[
\boldsymbol{V_{H}}=\begin{bmatrix}\left(\mu_{H}+\gamma_{H}\right) & 0\\
-\gamma_{H} & \mu_{H}
\end{bmatrix}\textrm{, }
\boldsymbol{V_{BB}} = \begin{bmatrix}\left(\left(\mu+\eta\right)\boldsymbol{I}-\boldsymbol{A}^{T}\right) & \boldsymbol{0}_{n\times n}\\
-\eta\boldsymbol{I} & \left(\mu\boldsymbol{I}-\boldsymbol{A}^{T}\right)
\end{bmatrix},
\]
\[
\boldsymbol{V_{RB}}=\begin{bmatrix}\boldsymbol{0}_{n\times1} & \boldsymbol{0}_{n\times1} & -\gamma_{R}\vec{\alpha} & \boldsymbol{0}_{n\times1}\\
\boldsymbol{0}_{n\times1} & \boldsymbol{0}_{n\times1} & \boldsymbol{0}_{n\times1} & -\gamma_{R}\vec{\alpha}
\end{bmatrix}\textrm{, }
\boldsymbol{V_{BV}} = \begin{bmatrix}-\left(-\boldsymbol{A}\vec{1}\right)^{T} & \boldsymbol{0}_{1\times n}\\
\boldsymbol{0}_{1\times n} & -\left(-\boldsymbol{A}\vec{1}\right)^{T}\\
\boldsymbol{0}_{1\times n} & \boldsymbol{0}_{1\times n}\\
\boldsymbol{0}_{1\times n} & \boldsymbol{0}_{1\times n}
\end{bmatrix},
\]
\[
\boldsymbol{V_{VR}}=\begin{bmatrix}\left(\mu+\gamma_{V}+\eta\right) & 0 & 0 & 0\\
-\eta & \left(\mu+\gamma_{V}\right) & 0 & 0\\
-\gamma_{V} & 0 & \left(\mu+\gamma_{R}+\eta\right) & 0\\
0 & -\gamma_{V} & -\eta & \left(\mu+\gamma_{R}\right)
\end{bmatrix}.
\]

The next-generation matrix is $\boldsymbol{K}=\boldsymbol{F}\boldsymbol{V}^-1$. 
The inverse of $\boldsymbol{V}$ can be formally written in block-matrix form as
\[
\boldsymbol{V}^{-1}=\begin{bmatrix}\boldsymbol{V_{H}}^{-1} & \boldsymbol{0}_{2\times2n+2} & \boldsymbol{0}_{2\times2}\\
\boldsymbol{0}_{2n+2\times2} & \boldsymbol{P} & -\boldsymbol{P}\boldsymbol{V_{RB}}\boldsymbol{V_{VR}}^{-1}\\
\boldsymbol{0}_{2\times2} & -\boldsymbol{V_{VR}}^{-1}\boldsymbol{V_{BV}}\boldsymbol{P} & \boldsymbol{V_{VR}}^{-1}+\boldsymbol{V_{VR}}^{-1}\boldsymbol{V_{BV}}\boldsymbol{P}\boldsymbol{V_{RB}}\boldsymbol{V_{VR}}^{-1}
\end{bmatrix}.
\]
where we used the substitution $\boldsymbol{P} = \left(\boldsymbol{V_{BB}}-\boldsymbol{V_{RB}}\boldsymbol{V_{VR}}^{-1}\boldsymbol{V_{BV}}\right)^{-1}$.
Therefore, the next-generation matrix is
\[
\boldsymbol{K}=\begin{bmatrix}\boldsymbol{0}_{2\times2} & \boldsymbol{F_{VH}}\boldsymbol{P} & -\boldsymbol{F_{VH}}\boldsymbol{P}\boldsymbol{V_{RB}}\boldsymbol{V_{VR}}^{-1}\\
\boldsymbol{F_{HV}}\boldsymbol{V_{H}}^{-1} & \boldsymbol{0}_{2n\times2n} & \boldsymbol{0}_{2n+2\times4}\\
\boldsymbol{0}_{4\times2} & \boldsymbol{0}_{4\times2n} & \boldsymbol{0}_{4\times4}
\end{bmatrix}
\]
We can now calculate the eigenvalues of $\boldsymbol{K}$,
\begin{align*}
 \left|\lambda\boldsymbol{I}-\boldsymbol{K}\right|&=\begin{vmatrix}\lambda\boldsymbol{I}_{2} & -\boldsymbol{F_{VH}}\boldsymbol{P} & \boldsymbol{F_{VH}}\boldsymbol{P}\boldsymbol{V_{RB}}\boldsymbol{V_{VR}}^{-1}\\
-\boldsymbol{F_{HV}}\boldsymbol{V_{H}}^{-1} & \lambda\boldsymbol{I}_{2n+2} & \boldsymbol{0}_{2n+2\times2}\\
\boldsymbol{0}_{2\times2} & \boldsymbol{0}_{2\times2n+2} & \lambda\boldsymbol{I}_{2}
\end{vmatrix},\\\left|\lambda\boldsymbol{I}-\boldsymbol{K}\right|&=\lambda^{2}\begin{vmatrix}\lambda\boldsymbol{I}_{2} & -\boldsymbol{F_{VH}}\boldsymbol{P}\\
-\boldsymbol{F_{HV}}\boldsymbol{V_{H}}^{-1} & \lambda\boldsymbol{I}_{2n+2}
\end{vmatrix},\\\left|\lambda\boldsymbol{I}-\boldsymbol{K}\right|&=\lambda^{2n+2}\det\left(\lambda^{2}\boldsymbol{I}_{2}-\boldsymbol{F_{VH}}\boldsymbol{P}\boldsymbol{F_{HV}}\boldsymbol{V_{H}}^{-1}\right).
\end{align*}
Therefore the eigenvalues of $\boldsymbol{K}$ are zero (multiplicity $2n+2$) and the square root of the eigenvalues of $\boldsymbol{L}=\boldsymbol{F_{VH}}\boldsymbol{P}\boldsymbol{F_{HV}}\boldsymbol{V_{H}}^{-1}$. 

The $i,j$ entry of \mbox{$\boldsymbol{\Gamma_{I}}=[(\mu\boldsymbol{I}-\boldsymbol{A})-(\frac{\gamma_{V}}{\mu+\gamma_{V}})(\frac{\gamma_{R}}{\mu+\gamma_{R}})(-\boldsymbol{A}\vec{1})\vec{\alpha}^{T}]^{-1}$} gives the average amount of time spent infectious and in biting stage $j$ given initially becoming infectious in biting stage $i$. 
Similarly, the $i,j$ entry of the matrix \mbox{$\boldsymbol{\Gamma_{E}}=[((\mu+\eta)\boldsymbol{I}-\boldsymbol{A})-(\frac{\gamma_{V}}{\mu+\gamma_{V}+\eta})(\frac{\gamma_{R}}{\mu+\gamma_{R}+\eta})(-\boldsymbol{A}\vec{1})\vec{\alpha}^{T}]^{-1}$} gives the average amount of time spent in biting stage $j$ while exposed, given initial exposure in biting stage $i$. 
The $i,j$ entry of the matrix \mbox{$\boldsymbol{\tau_{E}}=\left(\eta\boldsymbol{I}+(-\boldsymbol{A}\vec{1})\left[(\frac{\eta}{\mu+\gamma_{V}+\eta})(\frac{\gamma_{V}}{\mu+\gamma_{V}})+(\frac{\gamma_{V}}{\mu+\gamma_{V}+\eta})(\frac{\eta}{\mu+\gamma_{R}+\eta})\right](\frac{\gamma_{R}}{\mu+\gamma_{R}})\vec{\alpha}^{T}\right)\boldsymbol{\Gamma_{E}}$} is the probability that a mosquito that became exposed in biting stage $i$ survives to become infectious in biting stage $j$.

With these definitions, the matrix $\boldsymbol{P} = \left(\boldsymbol{V_{BB}}-\boldsymbol{V_{RB}}\boldsymbol{V_{VR}}^{-1}\boldsymbol{V_{BV}}\right)^{-1}$ can be expressed as
\[
\left(\boldsymbol{V_{BB}}-\boldsymbol{V_{RB}}\boldsymbol{V}_{V}^{-1}\boldsymbol{V_{VR}}\right)^{-1}=\begin{bmatrix}\boldsymbol{\Gamma_{E}}^T & \boldsymbol{0}_{n\times n}\\
\boldsymbol{\Gamma_{I}}^T\boldsymbol{\tau_{E}}^T & \boldsymbol{\Gamma_{I}}^T
\end{bmatrix}.
\]

Hence 
\[
\boldsymbol{K}=\begin{bmatrix}\vec{1}^{T}\boldsymbol{\beta_{H}}\boldsymbol{\Lambda_{BH}}\boldsymbol{\Gamma_{I}}^T\boldsymbol{\tau_{E}}^T\boldsymbol{\beta_{B}}\boldsymbol{\Lambda_{HB}}\frac{\vec{B}^{*}}{B^{*}} & 0\\
0 & 0
\end{bmatrix}
\]
and the eigenvalues of $\boldsymbol{K}$ are zero (multiplicity $2n+4$) and $\pm\sqrt{(\frac{1}{\gamma_{H}+\mu_{H}})\vec{1}^{T}\boldsymbol{\beta_{H}}\boldsymbol{\Lambda_{BH}}\boldsymbol{\Gamma_{I}}^T\boldsymbol{\tau_{E}}^T\boldsymbol{\beta_{B}}\boldsymbol{\Lambda_{HB}}\frac{\vec{B}^{*}}{B^{*}}}$. Therefore, $\mathcal{R}_{0}=\rho(\boldsymbol{K})$ is
\[
\mathcal{R}_{0}=\sqrt{\left(\frac{1}{\gamma_{H}+\mu_{H}}\right)\vec{1}^{T}\boldsymbol{\beta_{H}}\boldsymbol{\Lambda_{BH}}\boldsymbol{\Gamma_{I}}^T\boldsymbol{\tau_{E}}^T\boldsymbol{\beta_{B}}\boldsymbol{\Lambda_{HB}}\frac{\vec{B}^{*}}{B^{*}}}.
\]
By Theorem 2 of \cite{van-den-Driessche2002-ck}, if $\Rzero{}<1$, the DFE is stable and if $\Rzero{}>1$, the DFE is unstable.

\subsection{Global sensitivity analysis: Extended Fourier Amplitude Sensitivity Test}\label{GSA:eFAST}
\begin{figure}[t!]
    \centering
    \includegraphics[width=\textwidth,trim={0.175in 0.06in 0.04in 0.05in},clip]{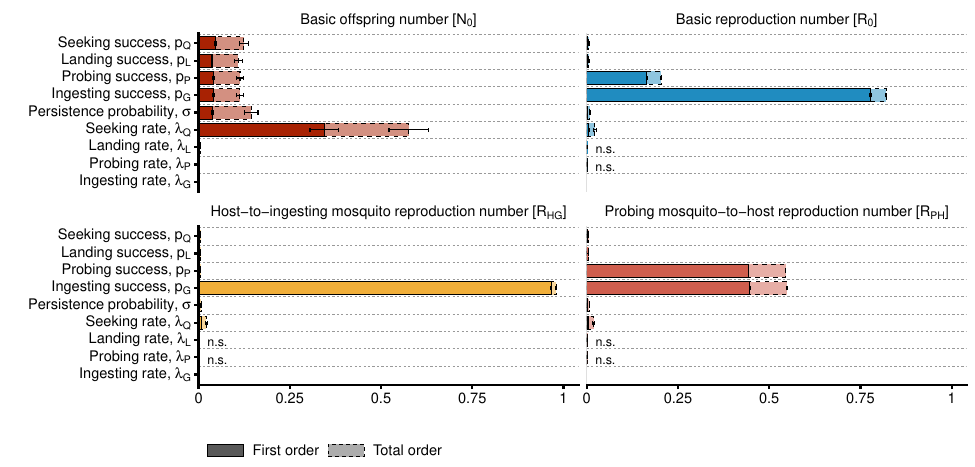}
    \caption{
    Extended Fourier amplitude sensitivity (eFAST) indices of the basic offspring number (red bars), basic reproduction number (blue bars), host-to-ingesting mosquito reproduction number (yellow bars), and probing mosquito-to-host reproduction number (pink bars) with respect to the mechanistic model biting parameters. 
    To avoid situations where the basic reproduction number was undefined (that is, when the basic offspring number is less than one), probabilities were sampled from a uniform distribution ranging from 20\% to 100\% except $\sigma$, which was varied from 1\% to 100\%. All rates were sampled from a uniform distribution ranging from once every thirty minutes to twice per minute, except for the questing rate, which varied from once every eight hours to once every nine minutes.
    The label ``n.s.'' indicates sensitivity indices that were not deemed significantly larger than those of a dummy parameter using a one-tailed Mann-Whitney test.
    }
    \label{fig:all_eFAST}
\end{figure}

To complement our LHS-PRCC global sensitivity analysis, we conducted an additional sensitivity analysis using the extended Fourier amplitude sensitivity test (eFAST). 
We used the same parameter ranges as in the LHS-PRCC analysis. 
All analysis was conducted in \textit{Julia} using the GlobalSensitivity package. 
Eleven harmonics were used in the Fourier series decomposition, and 11,000 samples were taken (reflecting one thousand samples for the ten parameters and a dummy parameter).
This process was repeated eleven times to obtain the mean and standard deviation of the total and first sensitivity indices of each parameter for each output variable. 
For each sensitivity index, a one-tailed Mann-Whitney test was used to determine whether it was significantly greater than the sensitivity index of a dummy parameter.

As in the LHS-PRCC analysis, variation in the basic offspring number, $\mathcal{N}_0$, was mostly due to the host-seeking rate ($\lambda_Q$). 
The success probability parameters have a similar level of influence on $\mathcal{N}_0$.  
The total order sensitivity indices were substantially larger than the first-order indices, indicating that non-linear effects had a strong influence on $\mathcal{N}_0$.

As in the LHS-PRCC analysis, the basic reproduction number, $\mathcal{R}_0$, was most strongly impacted by the probing and ingesting success probabilities. 
A closer look at the type reproduction numbers indicates that ingesting success has an extremely strong influence on the host-to-ingesting mosquito reproduction number and a mild effect on the probing mosquito-to-host reproduction number.
The probing and ingesting success probabilities have a roughly equal influence on the probing mosquito to host reproduction number. 
All of the rates except the host-seeking rate and ingesting rate had an insignificant effect on the reproduction numbers. 
Unlike for $\mathcal{N}_0$, first order effects explained the majority of the variation in the reproduction numbers.

\end{appendices}

\end{document}